\documentclass[twocolumn,letterpaper,aps,prc,superscriptaddress,longbibliography,nofootinbib,floatfix]{revtex4-2}

\usepackage{graphicx}	
\usepackage{multirow}

\usepackage{xspace}	

\newcommand{\snn}[1]{\mbox{$\sqrt{s_{_{NN}}}$ = #1\,GeV}}

\newcommand{\pt}{\mbox{$p_T$}\xspace}

\newcommand{\Araa}{\mbox{$\langle R_{xA} \rangle$}\xspace}

\newcommand{\raa}{\mbox{$R_{xA}$}\xspace}

\newcommand{\rxa}{\mbox{$R_{xA}$}\xspace}

\newcommand{\Npart}{\mbox{$N_{\rm part}$}\xspace}
\newcommand{\Ncoll}{\mbox{$N_{\rm coll}$}\xspace}
\newcommand{\Nproj}{\mbox{$N_{\rm proj}$}\xspace}

\newcommand{\sqs}{\mbox{$\sqrt{s}$}\xspace}
\newcommand{\sqsn}{\mbox{$\sqrt{s_{_{NN}}}$}\xspace}

\newcommand{\pp}{\mbox{$p$$+$$p$}\xspace}
\newcommand{\pA}{\mbox{$p$$+$A}\xspace}
\newcommand{\xA}{\mbox{$x$$+$A}\xspace}

\renewcommand{\AA}{\mbox{$A$$+$$A$}\xspace}
\newcommand{\AB}{\mbox{$x$$+$A}\xspace}

\newcommand{\dAu}{\mbox{$d$$+$Au}\xspace}

\newcommand{\pAu}{\mbox{$p$$+$Au}\xspace}
\newcommand{\pPb}{\mbox{$p$$+$Pb}\xspace}
\newcommand{\pAl}{\mbox{$p$$+$Al}\xspace}
\newcommand{\HeAu}{\mbox{$^{3}$He$+$Au}\xspace}

\newcommand{\AuAu}{\mbox{Au$+$Au}\xspace}

\newcommand{\piz}{\mbox{$\pi^0$}\xspace}

\newcommand{\gevc}{\mbox{GeV/$c$}\xspace}
\newcommand{\gevcc}{\mbox{GeV/$c^{2}$}\xspace}
\newcommand{\gev}{\mbox{GeV}\xspace}
\newcommand{\vtwo}{\mbox{$v_{\rm 2}$}\xspace}
\newcommand{\vthree}{\mbox{$v_{\rm 3}$}\xspace}

\def\fig#1{Fig.~\ref{#1}}
\def\figu#1{Figure~\ref{#1}}

\def\tab#1{Table~\ref{#1}}

\begin{document}

\title{Systematic study of nuclear effects in $p$$+$Al, $p$$+$Au, $d$$+$Au, and
$^{3}$He$+$Au collisions at $\sqrt{s_{_{NN}}}=200$ GeV using $\pi^0$
production}

\newcommand{\abilene}{Abilene Christian University, Abilene, Texas 79699, USA}
\newcommand{\augie}{Department of Physics, Augustana University, Sioux Falls, South Dakota 57197, USA}
\newcommand{\banaras}{Department of Physics, Banaras Hindu University, Varanasi 221005, India}
\newcommand{\barc}{Bhabha Atomic Research Centre, Bombay 400 085, India}
\newcommand{\baruch}{Baruch College, City University of New York, New York, New York, 10010 USA}
\newcommand{\bnlcoll}{Collider-Accelerator Department, Brookhaven National Laboratory, Upton, New York 11973-5000, USA}
\newcommand{\bnlphys}{Physics Department, Brookhaven National Laboratory, Upton, New York 11973-5000, USA}
\newcommand{\caucr}{University of California-Riverside, Riverside, California 92521, USA}
\newcommand{\charlesczech}{Charles University, Ovocn\'{y} trh 5, Praha 1, 116 36, Prague, Czech Republic}
\newcommand{\ciae}{Science and Technology on Nuclear Data Laboratory, China Institute of Atomic Energy, Beijing 102413, People's Republic of China}
\newcommand{\cns}{Center for Nuclear Study, Graduate School of Science, University of Tokyo, 7-3-1 Hongo, Bunkyo, Tokyo 113-0033, Japan}
\newcommand{\colorado}{University of Colorado, Boulder, Colorado 80309, USA}
\newcommand{\columbia}{Columbia University, New York, New York 10027 and Nevis Laboratories, Irvington, New York 10533, USA}
\newcommand{\czechtech}{Czech Technical University, Zikova 4, 166 36 Prague 6, Czech Republic}
\newcommand{\dapnia}{Dapnia, CEA Saclay, F-91191, Gif-sur-Yvette, France}
\newcommand{\debrecen}{Debrecen University, H-4010 Debrecen, Egyetem t{\'e}r 1, Hungary}
\newcommand{\elte}{ELTE, E{\"o}tv{\"o}s Lor{\'a}nd University, H-1117 Budapest, P{\'a}zm{\'a}ny P.~s.~1/A, Hungary}
\newcommand{\eszterhazy}{Eszterh\'azy K\'aroly University, K\'aroly R\'obert Campus, H-3200 Gy\"ongy\"os, M\'atrai \'ut 36, Hungary}
\newcommand{\ewha}{Ewha Womans University, Seoul 120-750, Korea}
\newcommand{\famu}{Florida A\&M University, Tallahassee, FL 32307, USA}
\newcommand{\fit}{Florida Institute of Technology, Melbourne, Florida 32901, USA}
\newcommand{\fsu}{Florida State University, Tallahassee, Florida 32306, USA}
\newcommand{\gsu}{Georgia State University, Atlanta, Georgia 30303, USA}
\newcommand{\hiroshima}{Hiroshima University, Kagamiyama, Higashi-Hiroshima 739-8526, Japan}
\newcommand{\howard}{Department of Physics and Astronomy, Howard University, Washington, DC 20059, USA}
\newcommand{\ihepprot}{IHEP Protvino, State Research Center of Russian Federation, Institute for High Energy Physics, Protvino, 142281, Russia}
\newcommand{\illuiuc}{University of Illinois at Urbana-Champaign, Urbana, Illinois 61801, USA}
\newcommand{\inrras}{Institute for Nuclear Research of the Russian Academy of Sciences, prospekt 60-letiya Oktyabrya 7a, Moscow 117312, Russia}
\newcommand{\instpasczech}{Institute of Physics, Academy of Sciences of the Czech Republic, Na Slovance 2, 182 21 Prague 8, Czech Republic}
\newcommand{\isu}{Iowa State University, Ames, Iowa 50011, USA}
\newcommand{\jaea}{Advanced Science Research Center, Japan Atomic Energy Agency, 2-4 Shirakata Shirane, Tokai-mura, Naka-gun, Ibaraki-ken 319-1195, Japan}
\newcommand{\jeonbuk}{Jeonbuk National University, Jeonju, 54896, Korea}
\newcommand{\jyvaskyla}{Helsinki Institute of Physics and University of Jyv{\"a}skyl{\"a}, P.O.Box 35, FI-40014 Jyv{\"a}skyl{\"a}, Finland}
\newcommand{\kek}{KEK, High Energy Accelerator Research Organization, Tsukuba, Ibaraki 305-0801, Japan}
\newcommand{\korea}{Korea University, Seoul 02841, Korea}
\newcommand{\kurchatov}{National Research Center ``Kurchatov Institute", Moscow, 123098 Russia}
\newcommand{\kyoto}{Kyoto University, Kyoto 606-8502, Japan}
\newcommand{\labllr}{Laboratoire Leprince-Ringuet, Ecole Polytechnique, CNRS-IN2P3, Route de Saclay, F-91128, Palaiseau, France}
\newcommand{\lahorelums}{Physics Department, Lahore University of Management Sciences, Lahore 54792, Pakistan}
\newcommand{\lawllnl}{Lawrence Livermore National Laboratory, Livermore, California 94550, USA}
\newcommand{\losalamos}{Los Alamos National Laboratory, Los Alamos, New Mexico 87545, USA}
\newcommand{\lpc}{LPC, Universit{\'e} Blaise Pascal, CNRS-IN2P3, Clermont-Fd, 63177 Aubiere Cedex, France}
\newcommand{\lund}{Department of Physics, Lund University, Box 118, SE-221 00 Lund, Sweden}
\newcommand{\lyon}{IPNL, CNRS/IN2P3, Univ Lyon, Université Lyon 1, F-69622, Villeurbanne, France}
\newcommand{\maryland}{University of Maryland, College Park, Maryland 20742, USA}
\newcommand{\mass}{Department of Physics, University of Massachusetts, Amherst, Massachusetts 01003-9337, USA}
\newcommand{\michigan}{Department of Physics, University of Michigan, Ann Arbor, Michigan 48109-1040, USA}
\newcommand{\muenster}{Institut f\"ur Kernphysik, University of M\"unster, D-48149 M\"unster, Germany}
\newcommand{\muhlenberg}{Muhlenberg College, Allentown, Pennsylvania 18104-5586, USA}
\newcommand{\myongji}{Myongji University, Yongin, Kyonggido 449-728, Korea}
\newcommand{\nagasaki}{Nagasaki Institute of Applied Science, Nagasaki-shi, Nagasaki 851-0193, Japan}
\newcommand{\nara}{Nara Women's University, Kita-uoya Nishi-machi Nara 630-8506, Japan}
\newcommand{\natmephi}{National Research Nuclear University, MEPhI, Moscow Engineering Physics Institute, Moscow, 115409, Russia}
\newcommand{\newmex}{University of New Mexico, Albuquerque, New Mexico 87131, USA}
\newcommand{\nmsu}{New Mexico State University, Las Cruces, New Mexico 88003, USA}
\newcommand{\northcg}{Physics and Astronomy Department, University of North Carolina at Greensboro, Greensboro, North Carolina 27412, USA}
\newcommand{\ohio}{Department of Physics and Astronomy, Ohio University, Athens, Ohio 45701, USA}
\newcommand{\ornl}{Oak Ridge National Laboratory, Oak Ridge, Tennessee 37831, USA}
\newcommand{\orsay}{IPN-Orsay, Univ.~Paris-Sud, CNRS/IN2P3, Universit\'e Paris-Saclay, BP1, F-91406, Orsay, France}
\newcommand{\peking}{Peking University, Beijing 100871, People's Republic of China}
\newcommand{\pnpi}{PNPI, Petersburg Nuclear Physics Institute, Gatchina, Leningrad region, 188300, Russia}
\newcommand{\pusan}{Pusan National University, Pusan 46241, Korea}
\newcommand{\riken}{RIKEN Nishina Center for Accelerator-Based Science, Wako, Saitama 351-0198, Japan}
\newcommand{\rikjrbrc}{RIKEN BNL Research Center, Brookhaven National Laboratory, Upton, New York 11973-5000, USA}
\newcommand{\rikkyo}{Physics Department, Rikkyo University, 3-34-1 Nishi-Ikebukuro, Toshima, Tokyo 171-8501, Japan}
\newcommand{\saispbstu}{Saint Petersburg State Polytechnic University, St.~Petersburg, 195251 Russia}
\newcommand{\saopaulo}{Universidade de S{\~a}o Paulo, Instituto de F\'{\i}sica, Caixa Postal 66318, S{\~a}o Paulo CEP05315-970, Brazil}
\newcommand{\seoulnat}{Department of Physics and Astronomy, Seoul National University, Seoul 151-742, Korea}
\newcommand{\stonybrkc}{Chemistry Department, Stony Brook University, SUNY, Stony Brook, New York 11794-3400, USA}
\newcommand{\stonycrkp}{Department of Physics and Astronomy, Stony Brook University, SUNY, Stony Brook, New York 11794-3800, USA}
\newcommand{\sungskku}{Sungkyunkwan University, Suwon, 440-746, Korea}
\newcommand{\tenn}{University of Tennessee, Knoxville, Tennessee 37996, USA}
\newcommand{\texsu}{Texas Southern University, Houston, TX 77004, USA}
\newcommand{\titech}{Department of Physics, Tokyo Institute of Technology, Oh-okayama, Meguro, Tokyo 152-8551, Japan}
\newcommand{\tsukuba}{Tomonaga Center for the History of the Universe, University of Tsukuba, Tsukuba, Ibaraki 305, Japan}
\newcommand{\vandy}{Vanderbilt University, Nashville, Tennessee 37235, USA}
\newcommand{\waseda}{Waseda University, Advanced Research Institute for Science and Engineering, 17  Kikui-cho, Shinjuku-ku, Tokyo 162-0044, Japan}
\newcommand{\weizmann}{Weizmann Institute, Rehovot 76100, Israel}
\newcommand{\wigner}{Institute for Particle and Nuclear Physics, Wigner Research Centre for Physics, Hungarian Academy of Sciences (Wigner RCP, RMKI) H-1525 Budapest 114, POBox 49, Budapest, Hungary}
\newcommand{\yonsei}{Yonsei University, IPAP, Seoul 120-749, Korea}
\newcommand{\zagreb}{Department of Physics, Faculty of Science, University of Zagreb, Bijeni\v{c}ka c.~32 HR-10002 Zagreb, Croatia}

\affiliation{\abilene}
\affiliation{\augie}
\affiliation{\banaras}
\affiliation{\barc}
\affiliation{\baruch}
\affiliation{\bnlcoll}
\affiliation{\bnlphys}
\affiliation{\caucr}
\affiliation{\charlesczech}
\affiliation{\ciae}
\affiliation{\cns}
\affiliation{\colorado}
\affiliation{\columbia}
\affiliation{\czechtech}
\affiliation{\dapnia}
\affiliation{\debrecen}
\affiliation{\elte}
\affiliation{\eszterhazy}
\affiliation{\ewha}
\affiliation{\famu}
\affiliation{\fit}
\affiliation{\fsu}
\affiliation{\gsu}
\affiliation{\hiroshima}
\affiliation{\howard}
\affiliation{\ihepprot}
\affiliation{\illuiuc}
\affiliation{\inrras}
\affiliation{\instpasczech}
\affiliation{\isu}
\affiliation{\jaea}
\affiliation{\jeonbuk}
\affiliation{\jyvaskyla}
\affiliation{\kek}
\affiliation{\korea}
\affiliation{\kurchatov}
\affiliation{\kyoto}
\affiliation{\labllr}
\affiliation{\lahorelums}
\affiliation{\lawllnl}
\affiliation{\losalamos}
\affiliation{\lpc}
\affiliation{\lund}
\affiliation{\lyon}
\affiliation{\maryland}
\affiliation{\mass}
\affiliation{\michigan}
\affiliation{\muenster}
\affiliation{\muhlenberg}
\affiliation{\myongji}
\affiliation{\nagasaki}
\affiliation{\nara}
\affiliation{\natmephi}
\affiliation{\newmex}
\affiliation{\nmsu}
\affiliation{\northcg}
\affiliation{\ohio}
\affiliation{\ornl}
\affiliation{\orsay}
\affiliation{\peking}
\affiliation{\pnpi}
\affiliation{\pusan}
\affiliation{\riken}
\affiliation{\rikjrbrc}
\affiliation{\rikkyo}
\affiliation{\saispbstu}
\affiliation{\saopaulo}
\affiliation{\seoulnat}
\affiliation{\stonybrkc}
\affiliation{\stonycrkp}
\affiliation{\sungskku}
\affiliation{\tenn}
\affiliation{\texsu}
\affiliation{\titech}
\affiliation{\tsukuba}
\affiliation{\vandy}
\affiliation{\waseda}
\affiliation{\weizmann}
\affiliation{\wigner}
\affiliation{\yonsei}
\affiliation{\zagreb}
\author{U.A.~Acharya} \affiliation{\gsu} 
\author{A.~Adare} \affiliation{\colorado} 
\author{C.~Aidala} \affiliation{\mass} \affiliation{\michigan} 
\author{N.N.~Ajitanand} \altaffiliation{Deceased} \affiliation{\stonybrkc} 
\author{Y.~Akiba} \email[PHENIX Spokesperson: ]{akiba@rcf.rhic.bnl.gov} \affiliation{\riken} \affiliation{\rikjrbrc} 
\author{H.~Al-Bataineh} \affiliation{\nmsu} 
\author{J.~Alexander} \affiliation{\stonybrkc} 
\author{M.~Alfred} \affiliation{\howard} 
\author{V.~Andrieux} \affiliation{\michigan} 
\author{A.~Angerami} \affiliation{\columbia} 
% ...
\collaboration{PHENIX Collaboration}  \noaffiliation

\date{\today}


\begin{abstract}


The PHENIX collaboration presents a systematic study of  {\bf inclusive} $\pi^0$ 
production from $p$$+$$p$, $p$$+$Al, $p$$+$Au, $d$$+$Au, and 
$^{3}$He$+$Au collisions at $\sqrt{s_{_{NN}}}=200$ GeV.  Measurements 
were performed with different centrality selections as well as the total 
inelastic, 0\%--100\%, selection for all collision systems.  For 
0\%--100\% collisions, the nuclear-modification factors, 
$R_{xA}$, are consistent with unity for $p_T$ above 8 GeV/$c$, 
but exhibit an enhancement in peripheral collisions and a suppression in 
central collisions.  The enhancement and suppression characteristics are 
similar for all systems for the same centrality class. It is shown that 
for high-$p_T$-$\pi^0$ production, the nucleons in the $d$ and $^3$He 
interact mostly independently with the Au nucleus and that the counter 
intuitive centrality dependence is likely due to a physical correlation 
between multiplicity and the presence of a hard scattering process. 
These observations disfavor models where parton energy loss has a 
significant contribution to nuclear modifications in small systems. 
Nuclear modifications at lower $p_T$ resemble the Cronin effect -- an 
increase followed by a peak in central or inelastic collisions and a 
plateau in peripheral collisions. The peak height has a characteristic 
ordering by system size as $p$$+$Au $>$ $d$$+$Au $>$ $^{3}$He$+$Au $>$ 
$p$$+$Al. For collisions with Au ions, current calculations based on 
initial state cold nuclear matter effects result in the opposite order, 
suggesting the presence of other contributions to nuclear modifications, 
in particular at lower $p_T$.

\end{abstract}

\maketitle

\section{Introduction} 

Measurements of transverse-momentum (\pt) distributions of particles 
produced in hadronic collisions are commonly used to obtain information 
from the interaction.  At the Relativistic Heavy Ion Collider (RHIC) at 
Brookhaven National Laboratory, studies of the 
nuclear-modification factor $R_{AA}$ of hadrons, defined as the ratio of 
the hadron yield per binary nucleon-nucleon collision in a given A$+$A 
system to the yield measured in \pp collisions, have led to significant 
insights. The discovery of the suppression of high \pt neutral pions and 
charged hadrons~\cite{ppg:003,STAR:2002ggv} in \AuAu collisions relative 
to scaled \pp collisions at the same energy, was one of the first hints 
of parton energy loss in the strongly coupled quark gluon plasma (QGP). 
The apparent absence of any suppression in reference spectra from \dAu 
collisions~\cite{ppg:028,STAR:2003pjh}, where the formation of QGP was 
not expected, was critical to establish parton energy loss as the origin 
of the observed suppression in \AuAu collisions. The subsequent 
systematic studies of the suppression pattern of \piz production in 
\AuAu collisions at \sqsn = 200\,GeV allowed for quantitative 
constraints on the medium transport 
coefficients~\cite{Adare:2008qa,STAR:2009fqa}.

Experimentally, evidence for cold-nuclear-matter effects was first 
observed in the late 1970s when the ratio of the production cross sections 
of hadrons from \pA to \pp was found to vary with 
\pt~\cite{Cronin:1974zm,Antreasyan:1978cw}. This variation was referred 
to as the ``Cronin effect": a suppression at low \pt followed by an 
enhancement around 2--5\,\gevc that vanishes towards larger \pt. 
Historically the Cronin effect was attributed to initial state hard 
scattering~\cite{Kuhn:1975dq,Lev:1983hh}, but this explanation remained 
unsatisfactory because it could not explain the much larger effect for 
protons compared to pions. Measurements of the momentum spectra at RHIC 
in the early 2000s renewed interest in the Cronin effect, and various 
theoretical models have been developed to explain it. Most models were 
based on hard and soft multiple scattering 
\cite{Wang:1998ww,Zhang:2001ce,Kopeliovich:2002yh,Accardi:2001ih,Vitev:2002pf}, 
but there were additional approaches involving gluon saturation 
\cite{Kharzeev:2003wz} or hadronization by quark recombination 
\cite{Hwa:2004zd}. To date, there is no full quantitative explanation of 
the Cronin effect.

There are striking similarities between long range particle correlations in 
\AA collisions and those observed in high multiplicity \pp and \pPb 
collisions at the Large Hadron Collider 
(LHC)~\cite{Khachatryan:2010gv,Abelev:2012ola,Aad:2012gla,CMS:2012qk}. This 
came as a surprise, because their presence in \AA collisions was typically 
associated with the collective expansion of the QGP.  Similar correlations 
were found in \dAu collisions at RHIC~\cite{Adare:2013piz}.  These findings 
have profound consequences for the interpretation of \pA collisions as a 
benchmark for cold-nuclear-matter effects and suggest that QGP could be 
produced in these systems.

The PHENIX experiment has used the versatility of RHIC, which allows for 
collisions of light nuclei, such as $p$, $d$, and $^3$He, with larger nuclei, 
for systematic studies of particle correlations in small systems. In all 
systems studied, high multiplicity events show large azimuthal 
anisotropies, measured as \vtwo and \vthree, that can be related to the 
initial geometry of the collision system and the build-up of collective 
behavior of the produced 
particles~\cite{Adare:2015ctn,Aidala:2016vgl,Aidala:2017ajz,Adare:2017wlc,PHENIX:2018lia}, 
which would be indicative of QGP formation.  This can also be seen at 
LHC energies where a measurement from \pPb collisions~\cite{Aad:2019ajj} 
shows \vtwo extending out past 20 \gevc in \pt.  These large azimuthal 
anisotropies also suggest the presence of radial flow in a hydrodynamic 
expansion, which would have an effect on the yield below a few \gevc.

Results from long range correlations have prompted great interest in 
finding other evidence of the possible formation of QGP in small 
systems, such as parton energy loss or thermal photon emission. In such 
studies, data sets are typically divided into ``centrality classes" 
according to the particle multiplicity measured at forward rapidity on 
the side of the outgoing larger nucleus~\cite{Adare:2013nff}. Indeed, in 
\pPb collisions at the LHC~\cite{ATLAS:2014cpa} and \dAu collisions at 
RHIC~\cite{Adare:2015gla}, a suppression of the jet yield at high \pt 
was found for central collisions. However, the same analyses show a 
significant enhancement of the jet yield in peripheral collisions, 
putting in question if the observed suppression is due to energy 
loss~\cite{Kang:2015mta} or whether there are other mechanisms at play, 
for example, $x$-dependent color fluctuation effects in 
protons~\cite{Alvioli:2014eda,Alvioli:2017wou} or biases in the 
centrality selection due to energy conservation~\cite{Kordell:2016njg}.

In this paper new data on the system size and centrality dependence of 
\piz production are presented over a wide \pt range from 1 to 20 \gevc 
from \pAl, \pAu, \dAu, and \HeAu collisions at \sqsn = 200\,GeV compared 
to \pp collisions at the same energy. The data samples were recorded by 
the PHENIX experiment at RHIC during 2008 (\pp $5.2~{\rm pb}^{-1}$, \dAu 
$80~{\rm nb}^{-1}$), 2014 (\HeAu $24~{\rm nb}^{-1}$), and 2015 (\pp 
$60~{\rm pb}^{-1}$, \pAl $0.5~{\rm pb}^{-1}$, \pAu $0.2~{\rm pb}^{-1}$). 
The new \pp data are combined with the published results from 
\pp data taking in 2005~\cite{Adare:2007dg}.

\section{Experimental Setup}

To reconstruct the \piz meson, the electromagnetic calorimeter (EMCal) 
in the central arms of the PHENIX detector is used. The EMCal is 
segmented into eight sectors, four in the west and four in the east arm 
of the PHENIX experiment. The sectors in each arm cover 90 degrees in 
azimuth and $\pm$0.35 in pseudorapidity. All sectors in the west and the 
two top sectors in the east arm are made of 2,592 lead-scintillator 
(PbSc) towers each. The other two sectors comprise lead-glass crystals. 
For the analyses presented here only the PbSc sectors were used. At a 
distance of 5 meters from the nominal interaction point the angular 
segmentation of the PbSc sectors is 
$\Delta\phi$~x~$\Delta\eta$~$\approx$~0.01~x~0.01. The energy resolution 
achieved is $\delta$$E$/$E$ $\approx$ 2.1\% $\oplus$ 8.3\%/$\sqrt{E[GeV]}$ 
and arrival times of clusters are recorded with a resolution of 
$\approx$0.5\,ns.  Further details can be found in 
Ref.~\cite{Aphecetche:2003zr}.

For event selection and for centrality characterization the beam-beam 
counters (BBCs) are used, one on the north and one on the south side of 
the central arms. For asymmetric collision systems, the smaller 
(projectile) nucleus travels towards the north side and the larger 
(target) nucleus travels towards the south side. Each BBC is comprised 
of 64 \v{C}erenkov counter modules. The BBCs are located at 
$\pm$1.44\,m from the interaction point and cover a pseudorapidity range of 
$3.0<|\eta|<3.9$.  The BBC modules have a timing resolution of $\approx$0.1\,ns.

While the EMCal and the BBC were identical for data taking in 2008, 
2014, and 2015, but there were new or modified detector components 
in each year. The most notable change was a silicon-vertex tracker 
(VTX) installed in the central-arm acceptance in 2011. Although the VTX 
and other new components are not used in this analysis, the effect on 
the material budget needs to be taken into account in 
the Geant3~\cite{Brun:1978fy} simulation used to calculate 
efficiency and acceptance corrections for each data set.

\section{Data Samples}

Several data samples were taken with different trigger conditions for 
each of the collision systems. The minimum-bias (MB) data samples 
require coincidental hits in each of the two BBCs. For the data recorded 
in 2014 and 2015 the event vertex was required to be within $\pm$10\,cm 
of the nominal z=0 position. For the data recorded in 2008 the 
requirement was $\pm$30\,cm.

The collected MB data samples correspond to $\approx$88\% of the inelastic 
cross section for \dAu and \HeAu, 84\% for \pAu, 72\% for \pAl, and 54\% 
for \pp. The events that are not recorded by the MB trigger involve 
mostly single diffractive (SD) nucleon-nucleon collisions, which 
predominantly produce particles at forward or backward rapidity and thus 
do not lead to coincident hits in both BBCs. As the number of binary 
nucleon-nucleon collisions (\Ncoll) increases from \pp to \HeAu 
collisions, the effect of an individual SD nucleon-nucleon collision is 
averaged out and a larger fraction of the inelastic cross section is 
captured by the MB trigger.

All MB data samples in the analysis, except for the \pp samples, are 
subdivided into four centrality classes using the charge measured in the 
south BBC. The selections are 0\%--20\%, 20\%--40\%, 40\%--60\%, and the 
remainder of the MB sample ($>$60\%). Here the percentage refers to the 
fraction of events relative to all inelastic collisions.

The high luminosity provided by RHIC enables the increase of the 
statistics at high \pt, beyond what the data acquisition bandwidth would 
allow using an MB trigger only, by taking data samples with a high 
energy threshold photon trigger, which PHENIX calls the ERT trigger. 
This trigger requires a minimum energy recorded in the EMCal  
segments (4x4 towers grouped to trigger tiles). Three different energy 
thresholds were used for each collision system. The ERT trigger 
thresholds are summarized in Table~\ref{tab:ertthreshold}. No 
coincidence in the BBC was required. These samples are again divided 
into the same centrality classes as the MB sample.

\begin{figure*}[htb]
\includegraphics[width=0.99\linewidth]{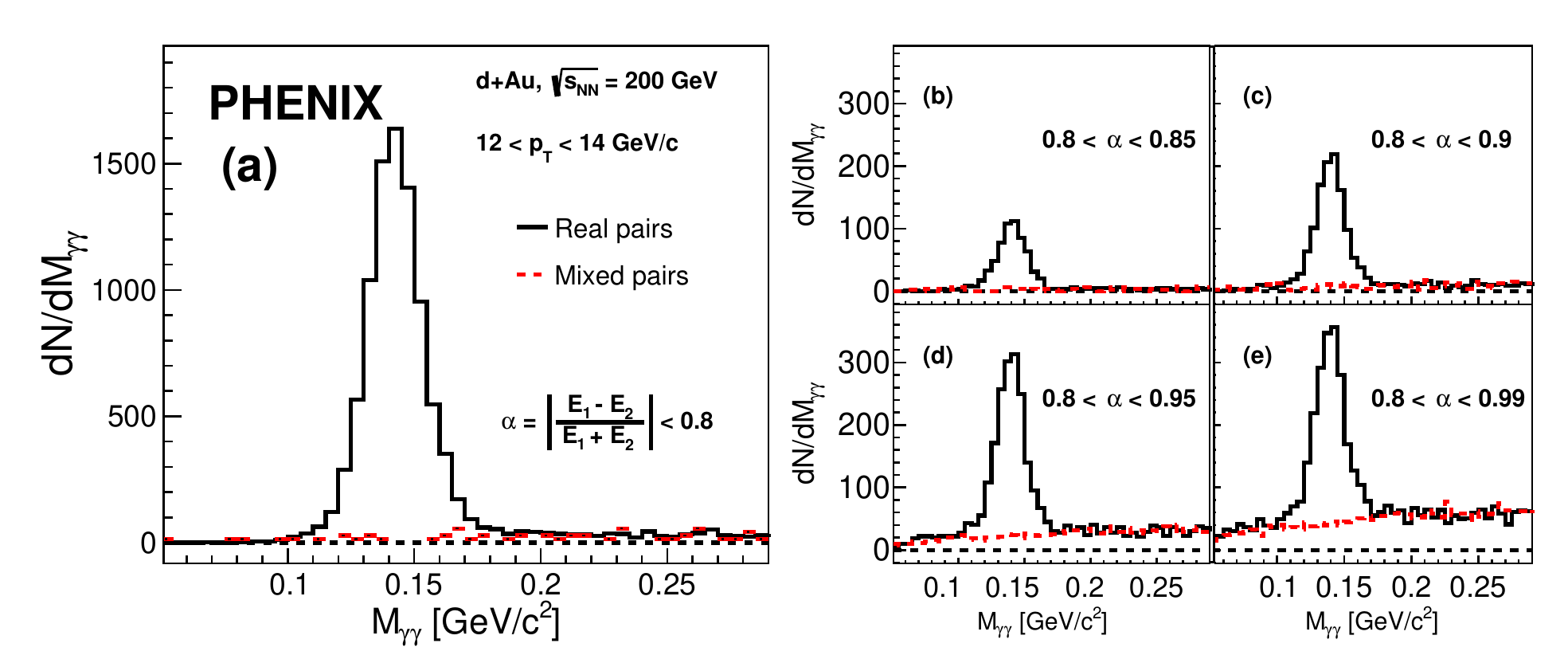}
\caption{(a) Invariant-mass example from \dAu collisions at 
$12<\pt<14\,\gevc$.  (b,c,d,e) The mass peak as a function of 
the asymmetry cut ($\alpha$) on the two photons for the indicated
$\alpha$ ranges.
}
\label{Fig:masslow}       
\end{figure*}

\begin{figure*}[htb]
\includegraphics[width=0.99\linewidth]{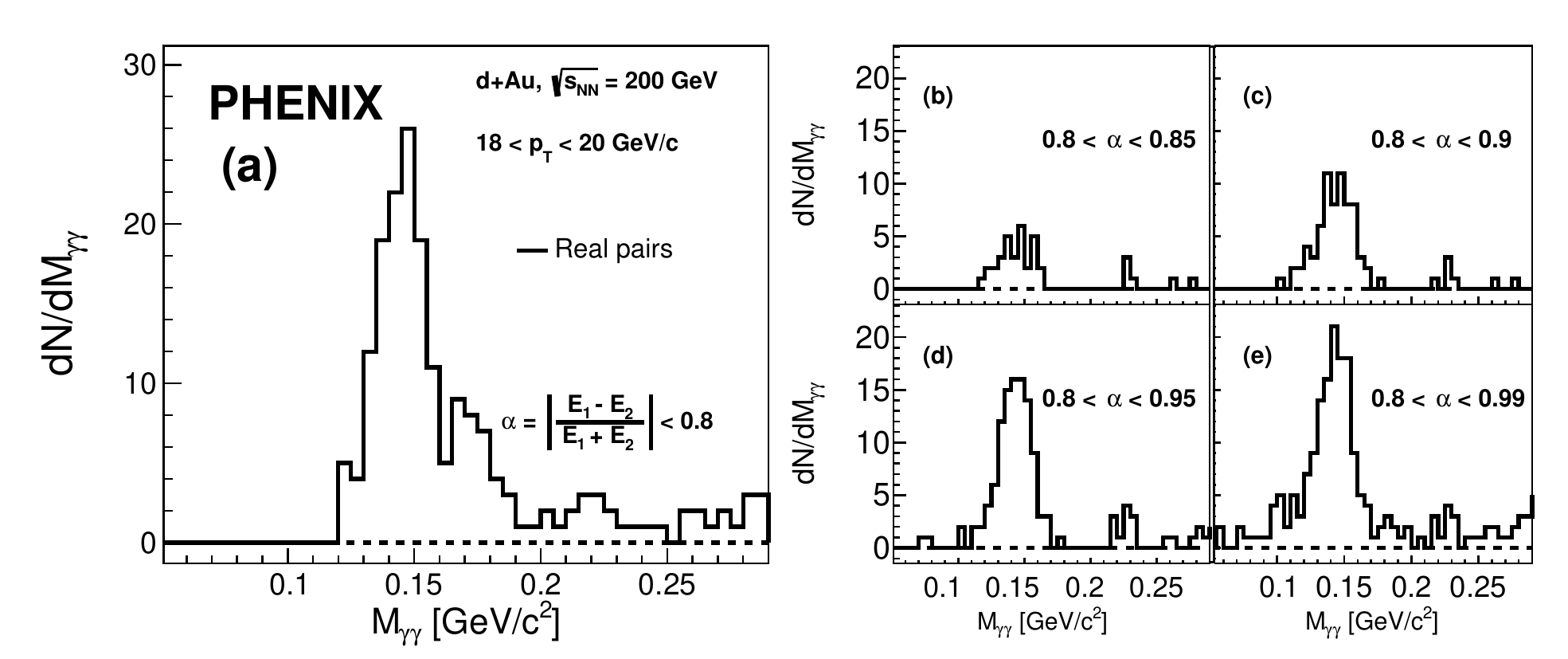}
\caption{(a) Invariant- mass example from \dAu collisions at 
$18<\pt<20\,\gevc$.  (b,c,d,e) The mass peak as a function of the 
asymmetry cut ($\alpha$) on the two photons for the indicated 
$\alpha$ ranges.
}
\label{Fig:masshigh}       
\end{figure*}

\begin{table}[htb]
    \caption{ERT trigger thresholds (GeV) for each collision system. 
}
    \begin{ruledtabular} \begin{tabular}{cccccc}
        &     \,\pp\, & \,\pAl\, & \,\pAu\, & \,\dAu\, & \,\HeAu \\
        \hline
        ERTA & 2.1 & 2.8 & 2.8 & 2.8  & 3.5 \\
        ERTB & 2.8 & 3.5 & 3.5 & 3.5  & 4.0 \\
        ERTC & 1.4 & 2.1 & 2.1 & 2.1  & 2.8 
    \end{tabular} \end{ruledtabular}
       \label{tab:ertthreshold}
\end{table}

During the \HeAu, \pAu, and \pAl data collection samples were also taken 
with a high multiplicity trigger. This trigger required, in addition to 
the BBC coincidence, a larger minimum charge in the south BBC, which 
corresponds to a larger number of fired BBC modules. The threshold was 
set to 25, 35, and 48 BBC modules, for \pAl, \pAu, and \HeAu 
respectively. The thresholds were chosen such that the data samples 
approximately correspond to the top 5\% most central collisions for each 
system.

\section{Data Analysis}

\subsection{Yield measurement}

Due to the high beam luminosity achieved at RHIC since 2010, PHENIX 
has recorded an increased number of double interactions that
are largest for the \pp data taken in 2015 and are  
noticeable for \pAu and \pAl data taken the same year.  The effect is 
negligible for the \pp, \dAu, and \HeAu data taken in 2008 and 2014, 
respectively. For the 2015 data, double interactions were reduced by 
making cuts on the time of flight measured for towers in the EMCal and 
the BBC modules. The cut on the EMCal requires the tower time to be 
within $\pm$5\,ns of the expected arrival time. This eliminates towers 
that are from different beam crossings. The BBC timing cut is used to 
reduce pile-up collisions that happen during the same bunch crossing.  
Such events are identified by large deviations of the time measured for 
individual BBC modules from the event average. For data from 2014 and 
2008 no cuts were applied. Any residual pileup events are accounted for 
in the systematic uncertainties.

The reconstruction of neutral pions is performed via the \piz 
$\rightarrow$ $\gamma\gamma$ decay channel.  The methods used by PHENIX 
have been described extensively in Ref.~\cite{Adler:2006bw} and will 
only be summarized in this paper. As a first step, neighboring PbSc 
towers with energy deposits above 0.015\,GeV are grouped into clusters. 
All clusters within one sector that have an energy of at least 0.3\,GeV 
are combined into pairs. A minimum distance of 8\,cm between the two 
cluster centers is required, corresponding to $\approx$1.5 tower separation 
between clusters. For each remaining pair, the invariant mass 
($M_{\gamma\gamma}$) and \pt are calculated. 
Invariant-mass distributions are generated in bins of \pt and collision 
centrality. All mass distributions show a clear peak at the \piz mass 
and a combinatorial background that is largest at events with low \pt 
and in central collisions.

To extract the \piz yield, the background in the \piz peak region needs 
to be subtracted. For \pt below $12\,\gevc$ an asymmetry cut of 
$\alpha<0.8$ is applied to reduce the combinatorial background. Here the 
asymmetry is defined as 
$\alpha =\left| \frac{(E_1-E_2)}{(E_1+E_2)}\right|$, where $E_1$ and 
$E_2$ are the energies of the two photon clusters. For \pt above 
12\,GeV/$c$ the cut is relaxed to $\alpha<0.95$ as discussed below.

The bulk of the background is estimated and subtracted by an event 
mixing technique that combines clusters from different events with 
similar vertex position ($z_{\rm vtx}$) and centrality. The shape of the 
mass distributions obtained from mixed events does not perfectly 
describe the combinatorial background in data. The mismatch results from 
correlated clusters in the event that are not accounted for in the mixed 
event technique.
 
For the MB samples, the mismatch is small and a two-step procedure is 
used for the subtraction. First, the mass distribution from mixed events 
is normalized in the mass region below and above the \piz peak, $0.05 < 
M_{\gamma\gamma} <0.1~\gevcc$ and $0.2<M_{\gamma\gamma}<0.4~\gevcc$, 
respectively. After subtracting the normalized distributions from all 
bins, a residual background remains. This is approximated by a {line} 
that is fitted to the same mass regions around the \piz peak and then 
also subtracted.
 
For the ERT data samples, the shape difference is more significant and 
thus a different approach is used. Instead of normalizing the mixed 
event distribution with a constant, the ratio of data/mixed events is 
fit with a second-order polynomial in the window around the \piz peak. 
This function is then used to normalize the mixed event distributions 
bin-by-bin, {in the same mass intervals below and above the mass peak as 
in the MB samples (see above)}. No residual background subtraction is 
needed in this case.
 
At very high-\pt, {typically larger than 15\,\gevc,} the combinatorial 
background is so small that neither normalization strategy for the mixed 
events gives stable results. Instead, the average count per mass bin, 
determined in the region below and above the \piz peak, is subtracted.
 
After the background subtraction, yields of \piz are calculated from 
the mass spectra by counting the entries within 2$\sigma$ of the peak, 
where the $\sigma$ is set by fitting the counts in the \piz region to a 
Gaussian.

\begin{figure}[htb]
\includegraphics[width=1.0\linewidth]{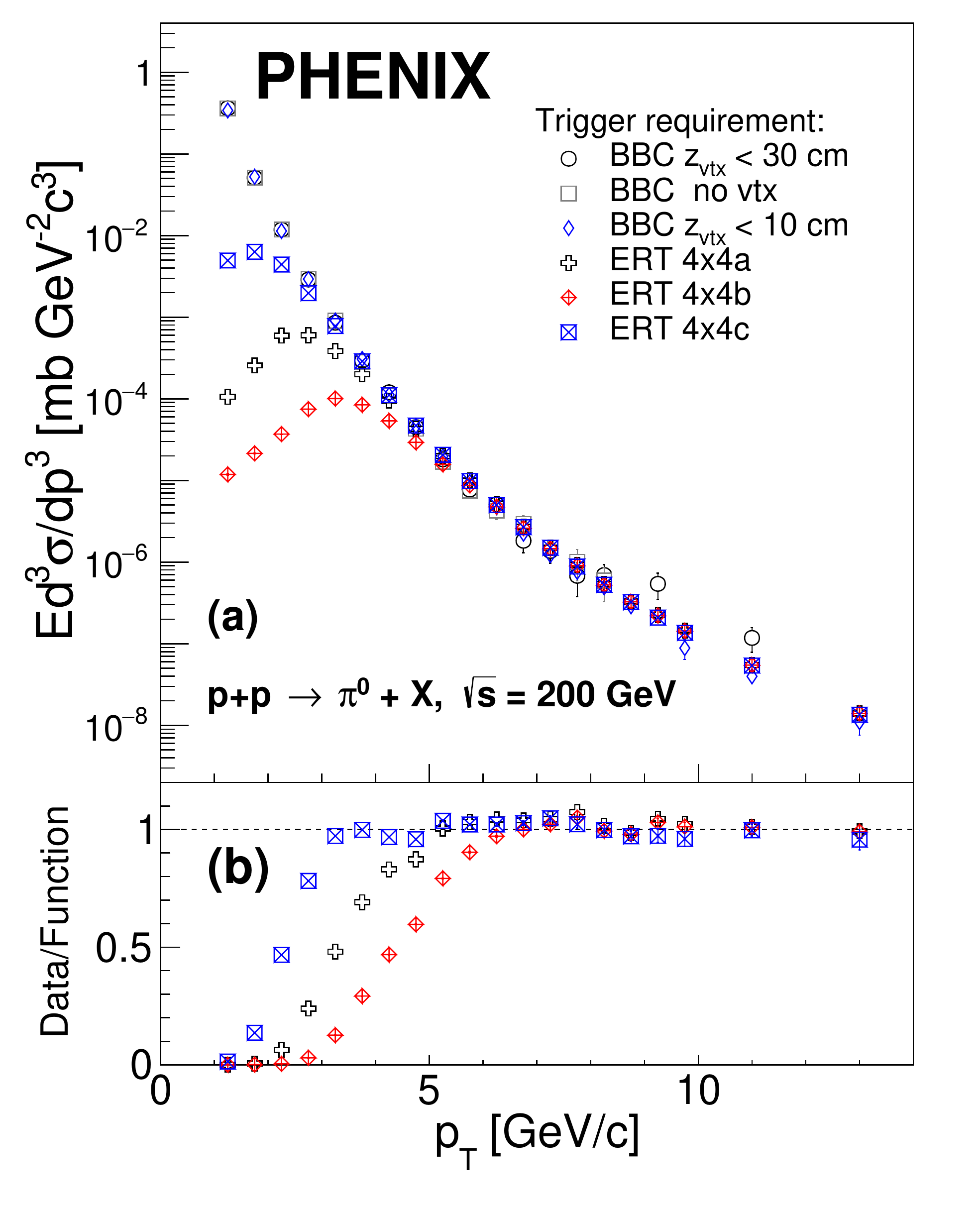}
\caption{(a) Invariant yield example from 2015 \pp collisions using 
different hardware trigger configurations. (b) The ratio of the 
different high-\pt triggers to a common Tsallis fit for all different 
triggers.
 }
\label{fig:ert}       
\end{figure}

\begin{figure}[htb]
\vspace{-0.5cm}
\includegraphics[width=1.0\linewidth]{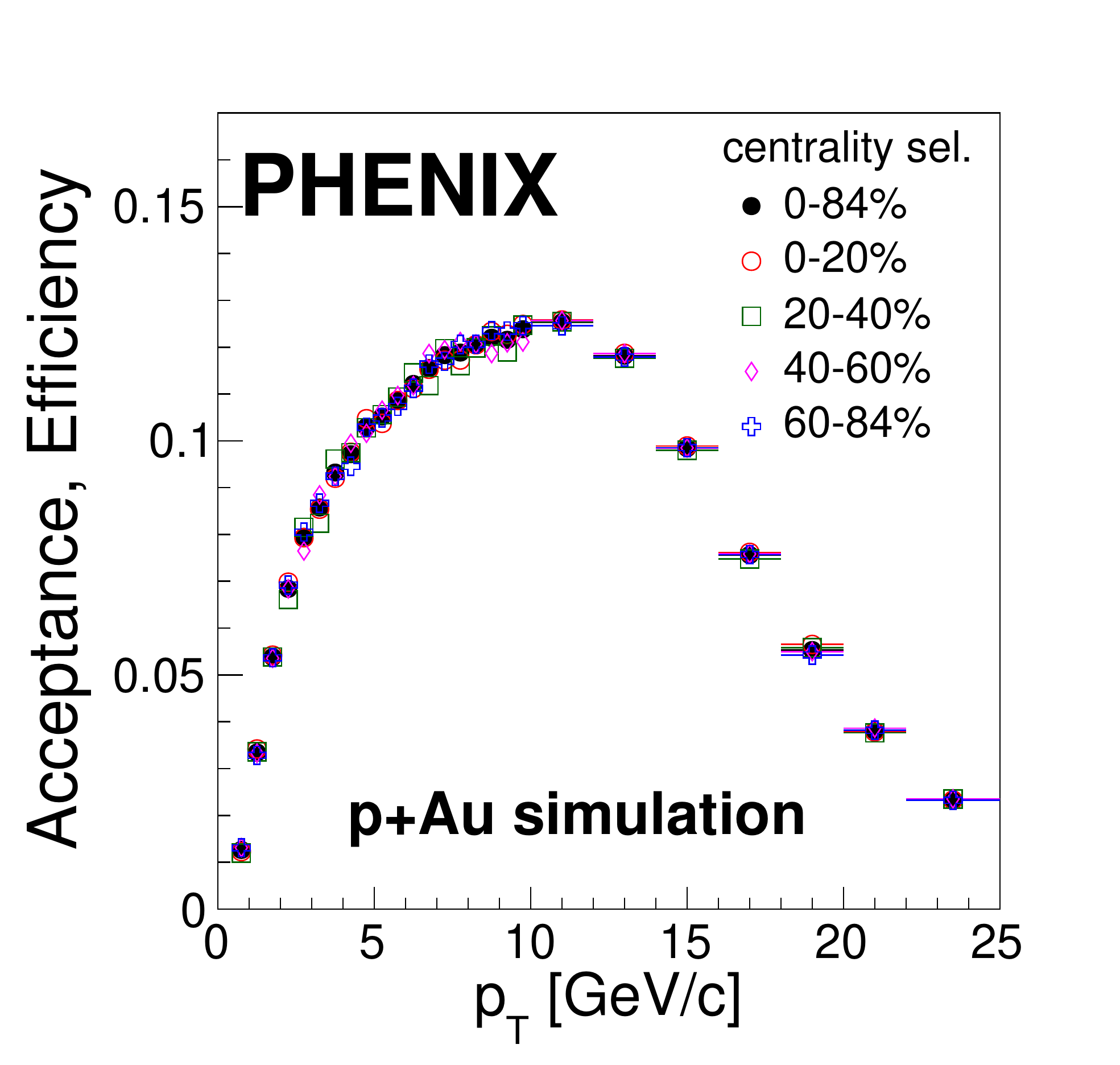}
\caption{The MC result of the acceptance and efficiency in \pAu 
collisions with the selected centrality classes as indicated.
 }
\label{Fig:corr}       
\end{figure}

\begin{table*}[htb]
\caption{Summary of the \Ncoll, \Npart, \Nproj, $f_{\rm bias}$ calculated 
using a Glauber MC simulation 
{\cite{Adare:2013nff,Adare:2018toe}}. The ratio $\Ncoll/\Nproj$ is also 
quoted for $d$ and $^3$He projectiles, because some systematic 
uncertainties cancel in this ratio. The last column is the measured 
charged particle multiplicity ($dN_{\rm ch}/d\eta$) in the midrapidity 
region {\cite{Adare:2018toe}}. \label{tab:Ncoll}
}
\begin{ruledtabular} \begin{tabular}{cccccccc}
system & centrality &$\langle \Ncoll	\rangle$ & $\langle \Npart	\rangle$ &
$\langle \Nproj	\rangle$ &$f_{\rm bias}$ & $\langle \Ncoll \rangle/\langle \Nproj \rangle$ & $dN_{\rm ch}/d\eta$ \\
\hline
\pp & & 1 & 2 & 1 & 0.73$\pm$0.07 & - & 2.38$\pm$0.09\\
\\
\pAl    &0\%--5\%      &4.1$\pm$0.3    &4.5$\pm$0.3    & 1     &0.81$\pm$0.01 &  - & 5.5$\pm$0.8\\
        &0\%--20\%	    &3.4$\pm$0.3	&4.4$\pm$0.3	& 1 	&0.81$\pm$0.01 &  - & 5.1$\pm$0.7\\
        &20\%--40\%	&2.3$\pm$0.1	&3.3$\pm$0.1	& 1	    &0.90$\pm$0.02 &  - & 4.0$\pm$0.6\\ 
        &40\%--60\%	&1.8$\pm$0.1 	&2.8$\pm$0.2	& 1	    &0.99$\pm$0.03 &  - & 3.3$\pm$0.3\\
        &60\%--72\%    &1.3$\pm$0.1  	&2.3$\pm$0.2	& 1 	&1.15$\pm$0.06 &  - & 2.7$\pm$0.1\\
        &0\%--100\%    &2.1$\pm$0.1    &3.1$\pm$0.1    & 1     &0.80$\pm$0.02  &  - & 4.0$\pm$0.5\\
\\
\pAu    &0\%--5\%      &9.7$\pm$0.6    &10.7$\pm$0.6   &1  &0.86$\pm$0.01  & - & 12.3$\pm$1.7\\
        &0\%--20\%     &8.2$\pm$0.5    &9.2$\pm$0.5    &1  &0.90$\pm$0.01  & - & 10.4$\pm$1.5\\
        &20\%--40\%    &6.1$\pm$0.4    &7.1$\pm$0.4    &1  &0.98$\pm$0.01  & - & 7.7$\pm$1.1\\
        &40\%--60\%    &4.4$\pm$0.3    &5.4$\pm$0.3    &1  &1.02$\pm$0.01  & - & 5.7$\pm$0.8\\
        &60\%--84\%    &2.6$\pm$0.2    &3.6$\pm$0.2    &1  &1.00$\pm$0.06  & - & 3.5$\pm$0.5\\
        &0\%--100\%    &4.7$\pm$0.3    &5.7$\pm$0.3    &1 &0.858$\pm$0.014 & - & 6.7$\pm$0.9\\
\\
\dAu    &0\%--5\%      &18.1$\pm$1.2   &17.8$\pm$1.2   &1.97$\pm$0.02 &0.91$\pm$0.01   & 8.98$\pm$0.59 & 18.9$\pm$1.4\\
        &0\%--20\%     &15.1$\pm$1.0    &15.2$\pm$0.6   &1.95$\pm$0.01 &0.94$\pm$0.01   & 7.46$\pm$0.50 & 16.4$\pm$1.2\\
        &20\%--40\%    &10.2$\pm$0.7   &11.1$\pm$0.6   &1.84$\pm$0.01 &1.00$\pm$0.01   & 5.71$\pm$0.39& 12.2$\pm$0.9\\
        &40\%--60\%    &6.6$\pm$0.4    &7.8$\pm$0.4    &1.65$\pm$0.02 &1.03$\pm$0.02   & 4.16$\pm$0.28 & 8.7$\pm$0.6\\
        &60\%--88\%    &3.2$\pm$0.2    &4.3$\pm$0.2    &1.36$\pm$0.02 &1.03$\pm$0.06   & 2.27$\pm$0.15 & 4.1$\pm$0.3\\
        &0\%--100\%    &7.6$\pm$0.4    &8.6$\pm$0.4    &1.62$\pm$0.01 &0.889$\pm$0.003 & 4.35$\pm$0.24 & 9.5$\pm$1.0\\
\\
\HeAu   &0\%--5\%      &26.1$\pm$2.0    &25.0$\pm$1.6   &2.99$\pm$0.01 &0.92$\pm$0.01  & 8.72$\pm$0.64 & 23.6$\pm$2.6\\
        &0\%--20\%     &22.3$\pm$1.7    &21.8$\pm$1.3   &2.95$\pm$0.01  &0.95$\pm$0.01 & 7.30$\pm$0.52 & 21.4$\pm$ 2.3\\
        &20\%--40\%    &14.8$\pm$1.1    &15.4$\pm$0.9   &2.75$\pm$0.03  &1.01$\pm$0.01 & 5.41$\pm$0.37 & 16.1$\pm$1.8\\
        &40\%--60\%    &8.4$\pm$0.6     &9.5$\pm$0.6    &2.29$\pm$0.04  &1.02$\pm$0.01 & 3.85$\pm$0.25 & 10.3$\pm$1.1\\
        &60\%--88\%    &3.4$\pm$0.3     &4.6$\pm$0.3    &1.56$\pm$0.05  &1.03$\pm$0.05 & 2.05$\pm$0.12 & 4.4$\pm$0.5\\
        &0\%--100\%    &10.4$\pm$0.7    &11.4$\pm$0.5   &2.22$\pm$0.02  &0.89$\pm$0.01 & 4.13$\pm$0.24 & 12.2$\pm$1.4\\
\end{tabular} \end{ruledtabular}
\end{table*}

Above $12~\gevc$, the two photon clusters from the \piz meson begin to 
{overlap more} and frequently merge into a single cluster. The asymmetry 
cut at $\alpha<0.8$, which was used to reduce the combinatorial 
background, starts to limit the \piz reconstruction efficiency and with 
it the effective \pt reach of the measurement. Because the combinatorial 
background is rather small at high \pt, the asymmetry cut can be relaxed 
to increase the reconstruction efficiency. Figures~\ref{Fig:masslow} 
and \ref{Fig:masshigh} show mass distributions from \dAu collisions in 
the $\pt = 12~{\rm to}~14\,\gevc$ and 18 to 20\,GeV/$c$ bins with 
different asymmetry cuts. The additional statistics recovered by 
extending the asymmetry {cuts are} clearly visible. In particular, in 
the higher \pt bin, increasing the cut from $\alpha<0.8$ to $<0.95$ 
effectively {increases} the statistics. Because it is also evident that 
the background increases, the looser cut is only used above 
$\pt>12~\gevc$. The background subtraction and \piz yield calculation 
follow the same steps as outlined above for lower \pt. The background 
estimate from event mixing is also shown on \fig{Fig:masslow}.  In 
\fig{Fig:masshigh}, the background is estimated from the average 
bin content around the \piz peak.

\subsection{Trigger selection}

At this stage of the analysis, raw \piz yields are available for all 
data samples in different bins of \pt and centrality. 
\figu{fig:ert}(a) compares the raw yields from the MB and ERT samples 
in \pp collisions from the 2015 data set. \figu{fig:ert}(b) shows the 
ratio of individual samples to a common fit. The ERT trigger turn on 
curves are clearly visible.

In the next step the raw yields from the MB and ERT trigger samples are 
combined for a given collision system and centrality. First, the ERT 
trigger samples are corrected for the trigger efficiency, {which is 
calculated as a function of the \piz \pt.  The trigger efficiency has a 
smooth turn on around} the trigger energy threshold and plateaus near 
100\% at higher \pt. A data driven method is used that compares the ERTC 
to the MB sample and the ERTA/ERTB to the ERTC sample to establish the 
turn on curve of the different trigger thresholds. The corrected spectra 
agree very well in the range where the trigger efficiency is larger than 
30\%.

To assure the largest statistical accuracy in each \pt bin, the MB 
triggered events are used in the low-\pt region, the ERTC trigger in the 
mid-\pt region, and the ERTB trigger at high-\pt. These transitions 
happen at different \pt thresholds for different collision systems. The 
\pt thresholds are set near the point where the trigger efficiency 
reaches its plateau value, typically close to twice the trigger 
threshold shown in \tab{tab:ertthreshold}. The ERTA triggered samples 
are used to crosscheck the results.

\subsection{Corrections to the yield}

Next, the raw \pt spectra need to be corrected for distortions due to 
the finite detector acceptance and overall detection efficiency 
(including detector effects and analysis cuts). These are determined 
simultaneously as one single correction as a function of \pt using a 
full Geant3 Monte Carlo (MC) simulation of the PHENIX detector setup. They 
are commonly referred to as acceptance-efficiency corrections 
(see Fig~\ref{Fig:corr}), which are determined separately for each 
centrality selection to account for any multiplicity dependent 
effects. For each running period, a separate simulation setup is used 
that describes the PHENIX detector configuration specific to that 
period. Samples of single \piz meson are simulated with a flat \pt 
distribution from 0 to 30~\gevc, full azimuthal coverage, and in one 
unit of rapidity at midrapidity. The resulting simulated detector 
responses are embedded into real data from the same running period and 
reconstructed using the same analysis methods applied to the data. The 
simulation was tuned so that \piz peak positions and widths 
reconstructed from the simulation matched the experimental data. Each 
reconstructed \piz is weighted with a realistic production probability 
for the \pt of the input \piz.  Because the true production probability is 
unknown, the weighting needs to be iterated. The probability is 
multiplied by the ratio of the measured raw \piz distribution over the 
reconstructed \piz distribution from the simulation. The modified 
probability is then used as the new weight. The process is iterated 
until convergence, which typically requires only a few steps. The final 
acceptance-efficiency corrections are calculated as the {ratio of the 
reconstructed number of \piz at a given \pt over the number of generated 
ones} at that \pt in one unit of pseudorapidity at midrapidity and 
2$\pi$ in azimuth.

Additionally, the yield in each centrality selection for a given 
collision system must be corrected for the bias towards higher event 
multiplicity, and hence {more central events,} for nondiffractive 
nucleon-nucleon collisions compared to diffractive collision events with 
the same impact parameter (see~\cite{Adare:2013nff} for full details). 
The bias factor $f_{\rm bias}$, which is used to scale the \pt spectra, is 
calculated using a Glauber Model MC 
calculation~\cite{Alver:2008aq} in conjunction with the assumption of a 
negative-binomial multiplicity distribution of particles produced in 
individual nucleon-nucleon collisions. The same Glauber calculation is 
used to characterize each centrality class by the number of binary 
nucleon-nucleon collisions \Ncoll, number of nucleon participants 
\Npart, and other relevant properties related to the collision geometry, 
such as \Nproj, the number of participants in the projectile nucleus. For 
MB collisions, the $f_{\rm bias}$ also includes the extrapolation from the 
recorded cross section to the full inelastic cross section (0\%--100\% 
centrality). The average values of \Ncoll, \Npart, \Nproj, and the bias 
factor $f_{\rm bias}$ are given in Table~\ref{tab:Ncoll}.

\section{Systematic Uncertainty}

\begin{table*}[htb]
\caption{\label{tab:Sys}
    Summary of systematic uncertainties on the \piz invariant yields 
from different running periods.
    }
 \begin{ruledtabular}   \begin{tabular}{cccccccccccc}
 Systematic uncer. & \multicolumn{3}{c}{2015 \pAu, \pAl, \pp} && \multicolumn{3}{c}{2014 \HeAu} && \multicolumn{3}{c}{2008 \dAu, \pp}\\ 
  \pt [\gevc]      & 2 & 8 & 20 & & 2 & 8 & 20 & & 2 & 8 &20 \\
  \hline
 Peak Extraction       & 4.4\% & 3.4\% & 1\%    && 2.7\%  & 4.1\%  & 2\%     & & 4.8\% & 2.9\% & 1.5\% \\
 Energy Scale          & 3.8\% & 6.5\% & 7.1\%  && 3.0\%  & 5.2\%  & 5.7\%   & & 4.6\% & 7.9\% & 8.7\% \\
 Acceptance-Efficiency & 3\%   & 2.5\% & 1\%    && 4\%     & 4\%    &  4\%   & & 3\%   & 2.5\% & 1\%   \\
 Cluster Merging       & $<$0.1\%   & $<$0.1\%  & 9.0\%  
                      && $<$0.1\%   & $<$0.1\%  & 12\%    
                      && $<$0.1\%   & $<$0.1\%  & 10\%  \\
 Conversion Loss       & 5\%   & 5\%   & 5\%    && 5\%    & 5\%    & 5\%     & & 2.5\% & 2.5\% & 2.5\% \\
 Double Interactions   & 4\%   & 3\%   & 4\%    && $<$1\% & $<$1\% & $<$1\%  & & 1\%   & 2.5\% & 4\%   \\
 Off Vertex Decays     & 3\%   & 3\%   & 3\%    && 3\%    & 3\%    & 3\%     & & 3\%   & 3\%   & 3\%   \\
 \\
 Total                 & 9.6\% & 10.1\% & 13.0\% && 8.3\%  & 9.8\%  & 14.1\%  && 8.3\% & 10.0\% & 14.5\%        
 \end{tabular}  \end{ruledtabular}
 \end{table*}

There are many sources of systematic uncertainty that need to be 
evaluated. They are separated into two groups: (i) uncertainty on the 
event characterization and (ii) uncertainty on the \piz yield 
extraction.

The event characterization is done using Glauber model simulations and 
the uncertainties were determined by varying the input to the Glauber 
model and various assumptions used in~\cite{Adare:2013nff}. The results 
are included in \tab{tab:Ncoll}. The quantities calculated from the 
Glauber model simulation are highly correlated.  For example, any change 
in the assumed nucleon-nucleon cross section will lead to a simultaneous 
change of \Ncoll, \Npart, and \Nproj. Thus, in ratios such as  
\Ncoll/\Nproj, some of the systematic uncertainties cancel. This was 
taken into account in the errors quoted in \tab{tab:Sys}.

The uncertainties on the \piz invariant yield are summarized in 
\tab{tab:Sys} for the different running periods. The total uncertainty 
on the \piz invariant yield varies between 8\%--10\% for \pt below 8 
\gevc and increases to nearly 15\% at 20 \gevc. They have little 
dependence on collision systems or centrality selection. The 
uncertainties on the \piz invariant yield were obtained with similar 
methods for all data sets. They are highly correlated within a running 
period and somewhat correlated between running periods. {In particular, 
the uncertainty on the energy scale and the uncertainty due to shower 
merging are correlated between all data sets. In 2014 and 2015, the 
experimental setup was nearly identical and therefore the 
acceptance-efficiency correction, losses due to photon conversions, and 
uncertainties due to off-vertex decays are also correlated for data sets 
taken during those years. For data sets taken within the same running 
period, all systematic uncertainties, except for the \piz peak 
extraction and the effect of double interactions, are correlated.} The 
correlations of the systematic uncertainties have been taken into 
account when combining data sets or calculating ratios of data sets { by 
determining the full error matrix and using the Best Linear Unbiased 
Estimate (BLUE) 
algorithm~\cite{Lyons:1988rp,Valassi:2003mu,Valassi:2013bga} to 
calculate the weight for each \pt and each measurement.}

The remainder of this section provides more details on the evaluation of 
the systematic uncertainties on the \piz yield determination, which is 
split into the extraction of the raw \piz yield and the corrections that 
need to be applied to it.

\subsection{Raw \piz yield extraction} 

The raw \piz yield is extracted from an invariant mass 
$M_{\gamma\gamma}$ distribution, which involves the subtraction of a 
background distribution below a \piz peak. Except for {at} very high 
\pt, this is done using the mixed event technique. This subtraction is 
typically accurate to better than 4\%. In general, the uncertainties on 
the background subtraction are determined by changing the assumption on 
the shape of the background and how it is normalized. Many different 
strategies can be used, as they all give similar results. Here, one 
example is given, the strategy that was used for the 2015 MB data sets, 
which were used to extract the \piz yield at lower \pt values for \pAu, 
\pAl, and \pp. The normalization of the mixed event background is 
determined in different ranges below and above the \piz peak. For any 
normalization, after the mixed event subtraction there is a residual 
background, which is then fitted. For each normalization the fit range 
is varied to extract the residual background via a first-order 
polynomial. Then in each case the window for the \piz yield extraction 
is varied from 1 to 3 sigma around the \piz peak. The variation of the 
resulting \piz yields, after correcting for the different $\sigma$ 
ranges, is used to estimate the systematic uncertainty.

The accuracy with which the \piz yield can be extracted depends on the 
amount of background.  In general, the smaller the particle multiplicity 
in the event and/or the larger the \piz \pt, the smaller the background. 
However, the accuracy with which the background can be determined for a 
particular \pt and centrality bin is driven by the available statistics. 
The dominant effect changes depending on the \piz \pt and the MB or ERT 
data set.

\subsection{Corrections of the raw yield}

The acceptance-efficiency correction accounts for all distortions to the 
\piz spectra that can be evaluated with the detailed simulation of \piz 
measurements in the PHENIX experiment. The accuracy of the simulation 
determines the size of systematic uncertainties. {Accordingly, the 
simulation output was carefully compared to the data.}

These distortions include, besides the actual corrections for detector 
acceptance and \piz reconstruction, the one for the energy scale and 
resolution, merging of clusters, and losses due to photon conversions. 
While the corrections were determined simultaneously, possible 
uncertainties are studied separately. In \tab{tab:Sys} these are 
identified as ``Acceptance-Efficiency", ``Energy Scale", ``Cluster 
Merging", and ``Conversion Loss", respectively.

The energy scale and resolution was tuned by matching the \piz peak 
position and width in simulation and data, as function of \pt, to a 
better than 0.5\%--1\% agreement, depending on the data set. The 
uncertainty is then determined by varying the energy scale and 
resolution within the achieved accuracy. The \piz yields {change by less 
than} 4\%--5\% at 2 \gevc and up to 7\%--9\% at 20 \gevc.

To study the accuracy of the reconstruction efficiency correction, cuts 
applied in the \piz reconstruction were varied and the analysis was 
repeated. The changes in the \piz yield were used to set the systematic 
uncertainties. They are typically smaller than 4\%, but may be limited 
by statistical uncertainties. The uncertainty on the acceptance was 
determined from the precision of the survey of the EMCal. It is 
negligible compared to the uncertainties on the reconstruction 
efficiency.

Because the two decay photons from the decay of a high \pt \piz are 
strongly boosted along the \piz direction, the average opening angle 
becomes small, resulting in only a small separation between the impact 
points on the surface of the EMCal. At $\approx$10 \gevc, the two clusters 
start to merge. Initially, this happens only for very symmetric decays 
characterized by a small energy asymmetry ($\alpha$). With increasing 
\pt, more and more clusters merge, leading to an {systematic decrease} 
in reconstruction efficiency towards higher \pt. The accuracy with which 
the MC simulation reproduces the cluster merging is verified by 
reconstructing \piz mesons from three exclusive asymmetry bins: 0--0.4, 
0.4--0.8, and 0.8--0.95. After fully correcting the \piz yields, the 
results are compared and the differences are used to estimate the 
systematic uncertainty. It reaches $\approx$10\% towards the end of the 
kinematic reach of the measurement.

\begin{figure}[htb]
\includegraphics[width=1.0\linewidth]{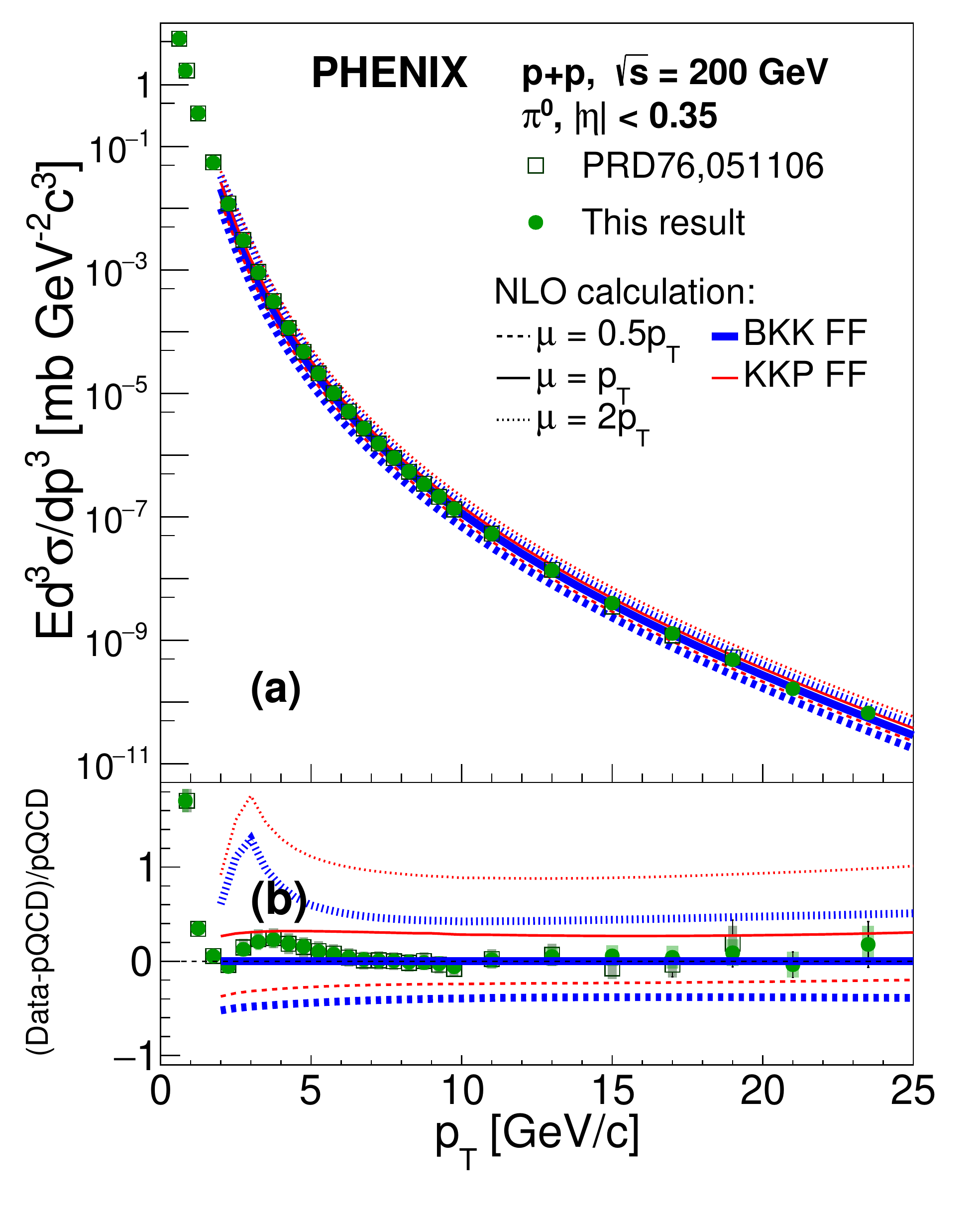}
\caption{(a) Differential cross section of \piz in \pp collisions at 
$\sqrt{s}=200$\,GeV. The data are compared with the indicated pQCD 
calculations. (b) The ratio of the data points to 
the NLO calculation with BKK and a scale of $\mu=\pt$.
}
\label{fig:crosssection}       
\end{figure}

Some photons convert into e$^+$e$^-$ pairs before they reach the EMCal. 
If the radial location of the conversion point is close to the EMCal, 
outside the magnetic field, the e$^+$ and $e^-$ will hit the EMCal in 
close proximity, resulting in one cluster with the full energy of the 
converted photon. In that case, it is likely that the \piz is 
reconstructed. However, if the conversion point is closer to the vertex, 
and in the magnetic field, the \piz will not be reconstructed, because the 
electron tracks bend in opposite direction{s}, depositing their energy 
in two separate clusters.

Prior to 2010, before the VTX was installed, $\approx$10\% of the \piz were 
not reconstructed because one of the photons converted in the detector 
material. Due to the additional material of the VTX detector close to 
the vertex, this number increases to $\approx$24\%. The accuracy with 
which the loss can be determined depends solely on the accuracy with 
which the material budget is known and implemented in the Geant3 
simulation. The resulting uncertainties on the \piz yield are 2.5\% and 
5\%, before and after installation of the VTX. There is no significant 
momentum dependence.

All data sets from 2015 (\pp, \pAl, and \pAu) were taken at high 
instantaneous luminosity, resulting in a significant number of 
recorded double interactions. These were actively identified and removed 
by timing cuts on the EMCal and BBC. The effect of any remnant double 
interaction was estimated by splitting the data samples into subsets 
taken at higher, medium, and lower luminosity. The analysis was repeated 
for each sample, and the \piz yields were found to be consistent within 
3\%--4\%. This difference was assigned as {a} systematic uncertainty. 
For the 2008 data sets (\pp and \dAu), only the EMCal timing cuts were 
applied to remove pileup events. Here, the possible contamination was 
estimated by the number of \piz for which at least one cluster had a 
time off by $>$5\,ns. The contribution was 1\% at high \pt and $\approx$4\% 
at lower \pt. For the 2014 \HeAu data no sizable effect was found.

Finally, the uncertainty of the normalization of the data taken with the 
ERT trigger to the MB data is examined. {It is estimated from the 
uncertainty on the linear fit of the ratio between the ERT and MB data 
in the region where the ERT trigger is fully efficient.}  This 
uncertainty is smaller than 1\% and not listed in \tab{tab:Sys}.

\begin{figure*}[bth]
\begin{minipage}[hbt]{0.99\linewidth}
\includegraphics[width=0.99\linewidth]{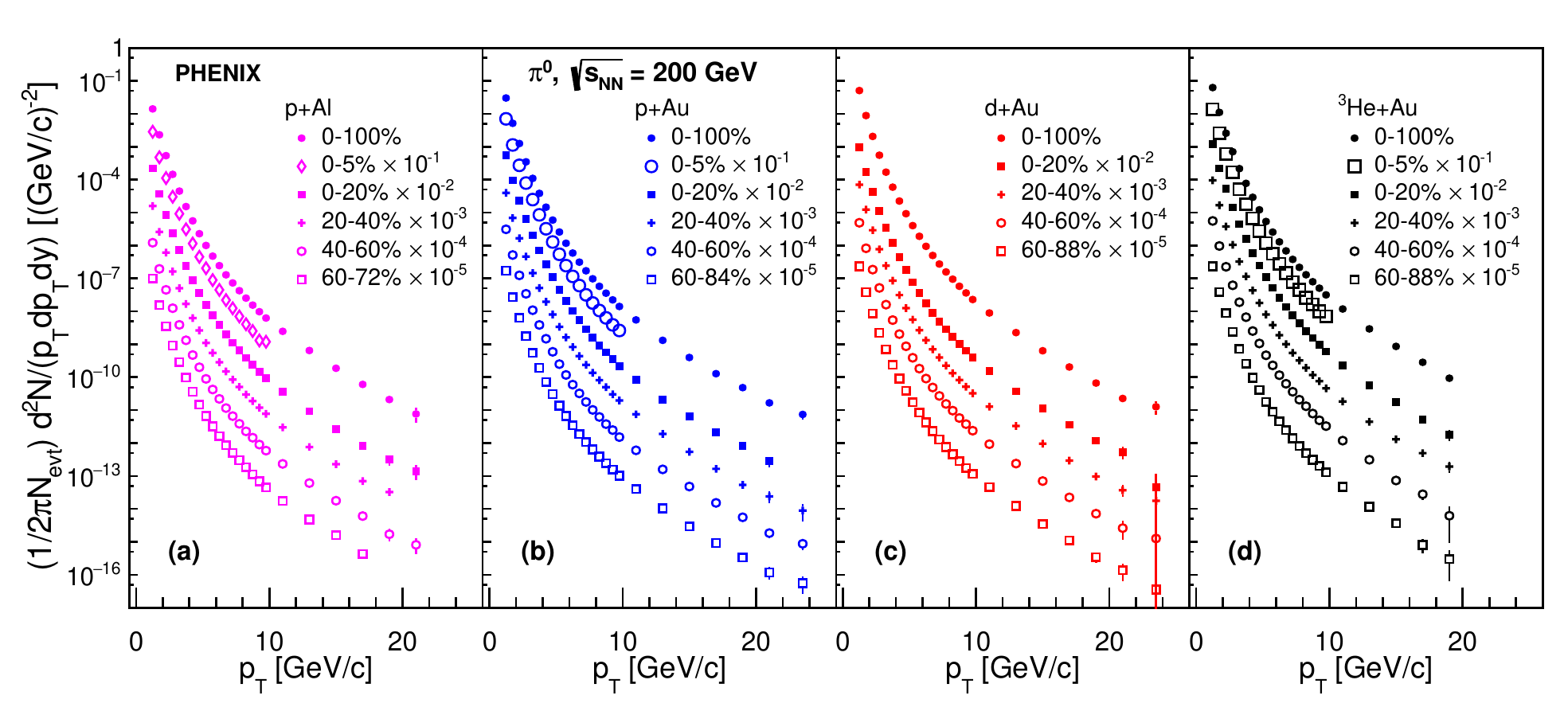}
\caption{Invariant yield of \piz from (a) \pAl, (b) \pAu, (c) \dAu, and 
(d) \HeAu at \sqsn = 200\,GeV. For each collision system the yield is 
shown for the inelastic cross section and for different centrality 
selections 0\%--20\%, 20\%--40\%, 40\%--60\%, and larger than 60\%. For 
\pAl, \pAu, and \HeAu an additional 0\%--5\% centrality selection is 
shown, which was recorded using a dedicated high multiplicity trigger.
}
\label{fig:yields}       
\end{minipage}
\begin{minipage}[hbt]{0.99\linewidth}
  \includegraphics[width=0.99\linewidth]{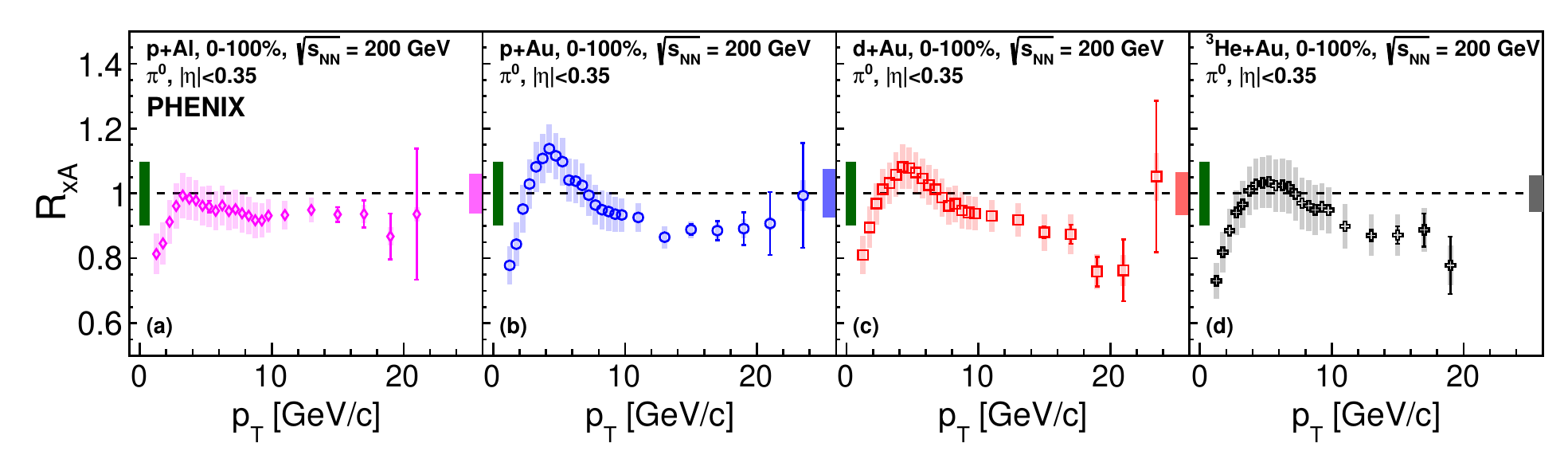}
\caption{Nuclear-modification factors from inelastic (a) \pAl, (b) \pAu, 
(c) \dAu, and (d) \HeAu collisions at $\sqsn = 200\,\gev$. The error 
bars represent the statistical uncertainties, while the boxes represent 
the systematic uncertainties. The high-\pt box in each panel is the 
\Ncoll uncertainty from the Glauber model, while the low-\pt box 
represents the overall normalization uncertainty from \pp collisions.
  }
  \label{fig:raamb}
\end{minipage}
\end{figure*}

\section{Results and discussion}

\subsection{The \pp reference}

PHENIX has previously published the \piz \pt spectrum from \pp 
collisions at $\sqs=200~\gev$~\cite{Adare:2007dg} based on data taken in 
2005 corresponding to $3.4~pb^{-1}$. In 2008 and 2015 RHIC provided 
further \pp collisions, increasing the integrated luminosity by 
$5.2~pb^{-1}$ and $60~pb^{-1}$ respectively.

With the {increase in the data sample}, the precision of the measurement 
was improved and extended to higher \pt. Because the detector 
configurations and the ERT trigger settings were different for the 2008 
and 2015 data sets, the \piz spectra were measured separately. The 
results were combined with those from 2005.

The new and published measurements were made with the PHENIX EMCal using 
the same analysis strategy, thus the \piz yield determinations have 
largely, but not completely, correlated systematic uncertainties. To 
combine the three data sets, the correlations between individual 
systematic uncertainties were studied and accounted for using the BLUE 
method~\cite{Valassi:2003mu}. In addition to the uncertainties due to 
the \piz reconstruction, there is an overall normalization uncertainty 
of 9.7\%~\cite{Adare:2007dg} that accounts for the limited accuracy with 
which the \pp MB trigger efficiency (see \tab{tab:Ncoll}) is known. This 
uncertainty is common to all \pp measurements.

\figu{fig:crosssection} compares the combined \piz \pt spectrum from \pp 
collisions (2005, 2008, 2015) to the earlier published 
result. The combined result is in excellent agreement with data taken in 
2005, but has significantly improved statistics and extends the \pt 
range up to 25\,\gevc. The systematic uncertainties are slightly reduced 
with respect to those of the 2005 data alone.

Also shown in \fig{fig:crosssection} are next-to-leading-order (NLO)  
perturbative-quantum-chromodynamics (pQCD)  
calculations~\cite{Jager:2002xm} with two different fragmentation 
functions (BKK and KKP) and for three different scales $\mu=\pt/2$, 
\pt, and 2\pt. For the calculations, the same CT14 free proton
parton distribution function (PDF) was used and only the 
fragmentation function in the same framework was changed. Within the 
assumed range of scales both fragmentation functions are consistent with 
the data. BKK would require a scale of $\mu=\pt$, while KKP envelopes 
the data between $\mu=\pt/2$ and \pt scales.

\subsection{Small system \pt spectra and nuclear-modification factor}

To simplify the labeling and description of each variable, the same 
notation is used for each small system.  The ``projectile" nucleus ($p$, 
$d$, or $^{3}$He) is denoted by $x$ and the ``target" nucleus (Au or Al) 
by A.  This notation is used in both the plots and text unless a 
specific system is being discussed.

\subsubsection{\pt spectra}

\figu{fig:yields} presents \piz \pt spectra from (a) \pAl, (b) \pAu, 
(c) \dAu, and (d)\HeAu. The data are presented as the 
invariant \piz yield per collision as a function of \pt. The 0\%--100\% 
range corresponds to the full inelastic cross section.  The other 
centrality ranges correspond to 0\%--5\%, 0\%--20\%, 20\%--40\%, 
40\%--60\%, and above 60\% measured percentile of the events selected 
according to the multiplicity measured in the BBC on the south side 
(heavy nucleus going side). Different centrality selections are scaled 
by factors 1/10 for visibility. The 0\%--5\% centrality selection, which 
is available for \HeAu, \pAu, and \pAl collisions, was taken with a high 
multiplicity BBC trigger and has a \pt range limited to below 10 \gevc.

\subsubsection{Nuclear-modification factor}

For a quantitative comparison across systems and centrality selections 
the nuclear-modification factor (\raa) is used. It is defined as:

\begin{equation}
\label{Eq:RAA}
\rxa = \frac{dN_{xA}/dp_{T} \times \sigma^{\rm inel}_{pp}}{\langle \Ncoll \rangle  \times  d\sigma_{pp}/dp_{T}} ,
\end{equation}

where $dN_{xA}/dp_{T}$ is the invariant yield per \xA collisions, 
$d\sigma_{pp}/dp_{T}$ is the invariant cross section in \pp collisions, 
$\sigma^{\rm inel}_{pp} = 42 {\rm mb}$ is the inelastic \pp cross section, 
and $\langle \Ncoll \rangle$ is the average number of binary 
nucleon-nucleon collisions given in \tab{tab:Ncoll}. The $\langle 
\Ncoll \rangle$ is obtained by the Glauber MC  
model~\cite{Miller:2007ri} used in all PHENIX papers and the 
detailed study of the model in smaller system centrality applications was described 
in~\cite{Adare:2013nff}. A nuclear-modification factor of 
$\rxa\approx1$ at high \pt indicates that \piz 
production through hard scattering processes in \xA collisions is well 
described by a{n incoherent} superposition of \pp collisions.

\subsubsection{\raa for inelastic collisions}

The nuclear-modification factors, \raa, for inclusive \piz production 
from inelastic \pAl, \pAu, \dAu, and \HeAu collisions are shown in 
\fig{fig:raamb}. They are calculated using the \pp reference from the 
combined 2005, 2008, and 2015 data. The correlations of the systematic 
uncertainties on the \piz reconstruction for different data sets are 
taken into account using the BLUE method. The overall {normalization} 
uncertainties on \pp and on \Ncoll are shown separately at the lowest
and highest \pt, respectively.

Each data set exhibits the characteristic \pt dependence of the Cronin 
effect, an initial rise from below unity to a peak around \pt of 
4\,\gevc, followed by a drop and a leveling off at high \pt. The 
constant value at high \pt is independent of the collision system at a 
value of \raa $\approx$ 0.9, which is consistent with unity within the 
systematic uncertainties on the scale and \Ncoll. The fact that \raa at 
high \pt is consistent with unity and that there is no system size 
dependence suggest that there is little to no modification of the hard 
scattering component in small systems.

To investigate any possible system size dependence of the modification 
at lower \pt, the ratio of the maximum of \raa divided by the integral 
taken above 10\,\gevc. This corresponds to the height of the peak in 
\raa assuming that \raa at high \pt is indeed unity. In these ratios the 
systematic uncertainties largely cancel. The values are $1.06\pm 0.09$, 
$1.25\pm 0.11$, $1.17\pm 0.10$, and $1.17\pm 0.12$ for \pAl, \pAu, \dAu, 
and \HeAu, respectively. The value is smallest in {\pAl collisions} and 
most pronounced in \pAu collisions. In addition, the maximum in \raa 
moves towards higher \pt with increasing system size from 3.25 \gevc in 
\pAl to 4.25 \gevc in \pAu and \dAu to 5.25 \gevc in $^3$He+Au.

The values are approximately the same as the peak heights calculated 
in fixed target \pA experiments~\cite{Straub:1992xd} and as originally 
predicted for RHIC 
energies~\cite{Wang:1998ww,Vitev:2002pf,Accardi:2001ih}.  However, the 
systematic trend with system size does not follow the dependence 
observed at fixed target energies~\cite{Antreasyan:1978cw},

\begin{equation}
\label{Eq:Cronin}
\frac{d\sigma_{xA}}{dp_T} = ({xA})^{n(p_T)} \times \frac{d\sigma_{pp}} {dp_T},
\end{equation}

with a common exponent $n(p_T)$ for a given $\sqrt{s}$. Here, $xA$ 
stands for the product of the number of nucleons in the small and large 
ions. Eq.~\ref{Eq:Cronin} is re-written in terms of per event yield by 
factoring out the inelastic cross sections $\sigma^{\rm inel}_{xA}$ 
and $\sigma_{pp}^{\rm inel}$:

\begin{equation}
\label{Eq:Cronin1}
\frac{dN_{xA}}{dp_{T}} =   (xA)^{n(p_{T})} \times  {\frac{\sigma^{\rm inel}_{pp}}{\sigma^{\rm inel}_{xA}} \times} \frac{dN_{pp}} {dp_{T}}.
\end{equation}

In the case of no nuclear modification for hard scattering processes, 
the per event yields in \xA and \pp collisions are related through the 
number of binary nucleon-nucleon collisions \Ncoll. In this case the 
exponent $n(\pt) =1$ and \Ncoll is:

\begin{equation}
\langle \Ncoll \rangle = {xA} \times \frac{\sigma^{\rm inel}_{pp}}{\sigma^{\rm inel}_{xA}}.
\end{equation}

This identity can be used to relate the nuclear-modification factor, 
\rxa, and the exponent $n(p_T)$:

\begin{equation}
\label{Eq:Croninexp}
n(p_T) = 1 + \frac{\log(R_{xA})} {\log(xA)} .
\end{equation}

\begin{figure}[htb]
\includegraphics[width=1.0\linewidth]{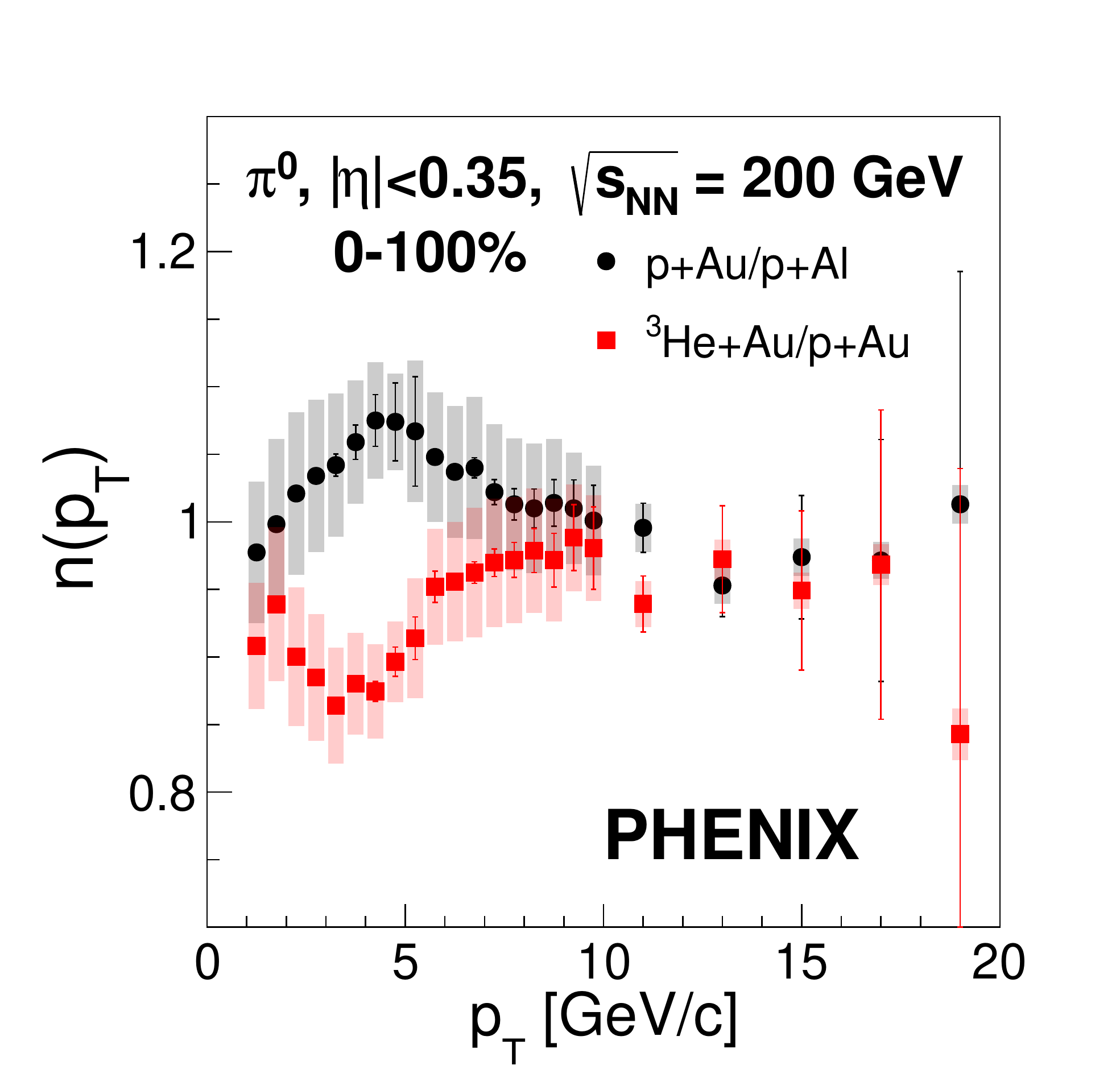}
\caption{Exponent according to Eq.~\ref{Eq:Croninexp} as a function 
of transverse momenta extracted from \pAu/\pAl and \HeAu/\pAu collision 
systems. The uncertainties from the \Ncoll calculations and from the 
overall normalization of \pp cancel in these ratios.
}
\label{fig:croninexp}       
\end{figure}

The exponent $n(p_T)$ is calculated from the ratio of 
$R_{p\rm{Au}}/R_{p\rm{Al}}$ and $R_{\rm{HeAu}}/R_{p\rm{Au}}$.  The 
uncertainties due to the \pp cross section cancel in the ratios; so do 
most of the uncertainties on the \Ncoll calculation.  The results are 
shown in \fig{fig:croninexp}. The data show that there is no universal 
$n(p_T)$ at \snn{200} below 8--10\,\gevc. At higher \pt, the common 
$n(p_T)$ underlines the similarity of \raa for all collision systems.

\subsubsection{\raa Centrality Dependence}

\begin{figure*}[htb]
\vspace{-1.0cm}
  \includegraphics[width=0.88\linewidth]{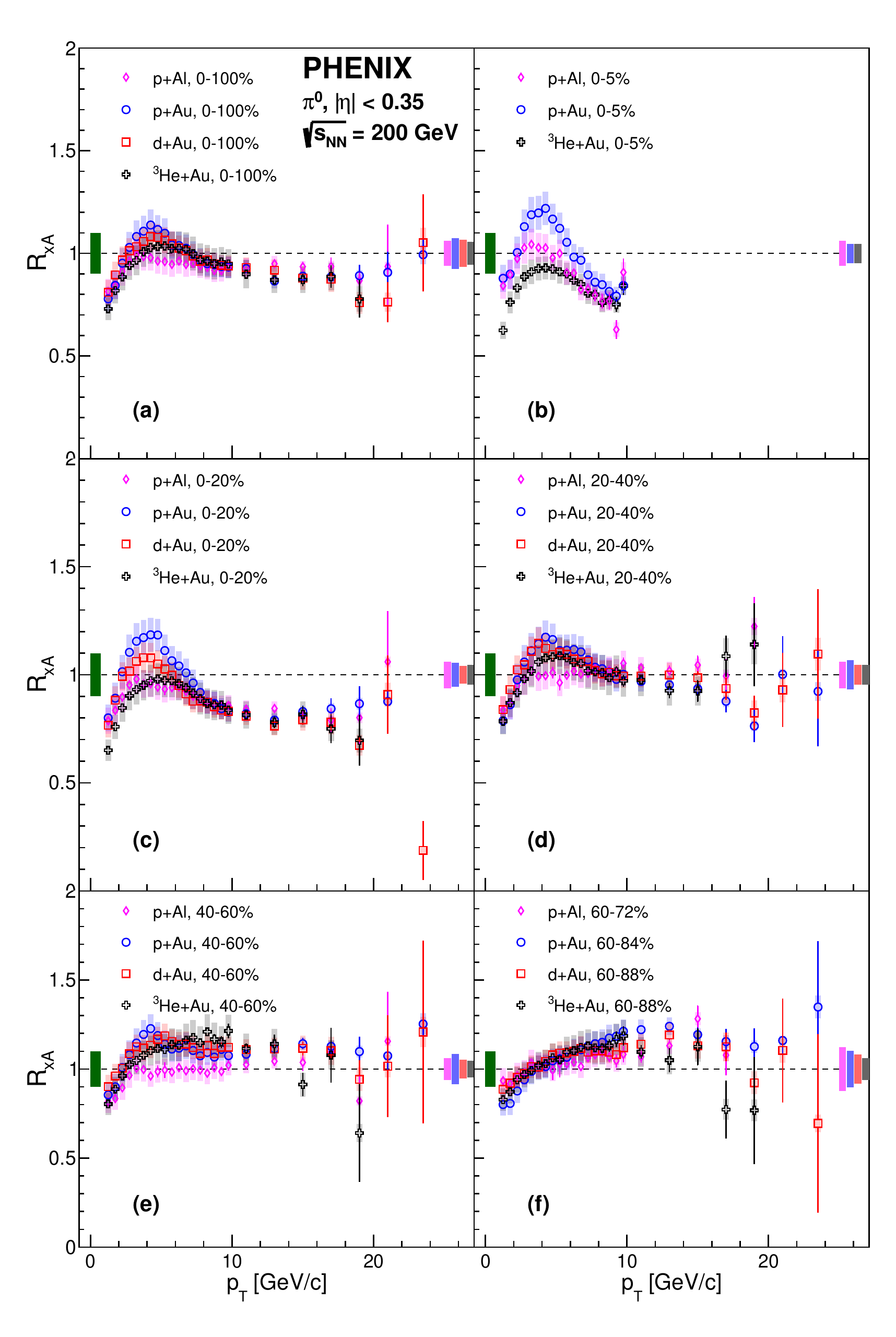}
  \caption{
Nuclear-modification factors in \pAl, \pAu, \dAu, and \HeAu in 0\%--100\% 
and the five indicated centrality bins and for inelastic collisions 
at $\sqsn=200$ GeV. The 
error bars represent the statistical uncertainties, while the boxes 
represent the systematic uncertainties. The high-\pt boxes are the 
uncertainties of the \Ncoll collisions from the Glauber model, while the 
low-\pt box represents the overall normalization uncertainty from 
\pp collisions.
}
  \label{fig:raa}
\end{figure*}

In Fig.~\ref{fig:raa}, \raa is shown for the different centrality 
selections from different collision systems. The scale uncertainty from 
the \pp reference and, to a large extent, the scale uncertainty due to 
\Ncoll only influences the scale of \raa, but not the relative 
differences between systems. The comparison reveals clear systematic 
trends of \raa with centrality and system size.

For $\pt>8\,\gevc$, the \raa values remain constant at similar values 
for the same centrality selection from different collision systems. 
However, the plateau value varies with centrality. \raa is below unity 
in the more central collisions, consistent with unity in the 20\%--40\% 
bin, and above or consistent with unity in the peripheral collisions. In 
the lower \pt range, the 0\%--5\% and 0\%--20\% selections exhibit a 
clear Cronin peak structure, similar to the {0\%--100\%} case, but more 
pronounced. The {height of the} peak is largest for \pAu. The height of 
the peak is system size dependent and decreases from \pAu, to \dAu, to 
\HeAu, i.e. with increasing size of the projectile nucleus. 
t The peak is smaller for \pAl than for \pAu, so it also seems to 
decrease with decreasing size of the target nucleus. In contrast, in 
peripheral collisions all systems follow a common trend. {Though there 
is a gradual change between central and semi-peripheral collisions, it 
is not consistent between systems.  }

To better understand the trends, the average nuclear-modification factor 
\Araa is calculated for two distinct \pt regions, above 8 \gevc to 
represent the high \pt region and from $4 < \pt < 6$ \gevc to capture 
the peak of \raa. These \Araa are studied as function of {\Ncoll and 
\Ncoll/\Nproj} shown in Tab. \ref{tab:Ncoll}. {Hard scattering processes 
are expected to scale with \Ncoll, thus \Ncoll is a natural choice. If 
the nucleons in the projectile interact independently with the target 
nucleus, nuclear modifications should not depend on \Ncoll, but rather 
\Ncoll/\Nproj.} Note that \Ncoll and \Npart are highly correlated and 
follow a {common} trend.  {In peripheral collisions, nucleons in the 
projectile are generally striking unique nucleons in the target and} 
$\Npart = \Ncoll +1$ up to an \Ncoll value of $\approx$14.  For \Ncoll $>$ 
14, \Npart increases slightly slower with \Ncoll as nucleons start to 
participate in multiple interactions.  Consequently, common trends of a 
nuclear modification with {\Ncoll} will also present themselves with 
respect {to \Npart}. The \Araa is calculated as follows:

\begin{equation}
    \Araa = \frac{\int \frac{dN_{xA}}{dp_T} dp_T}
                 {N_{\rm coll} \int \frac{dN_{pp}}{dp_T} dp_T}
\end{equation}

\begin{figure}[htb]
\includegraphics[width=1.0\linewidth]{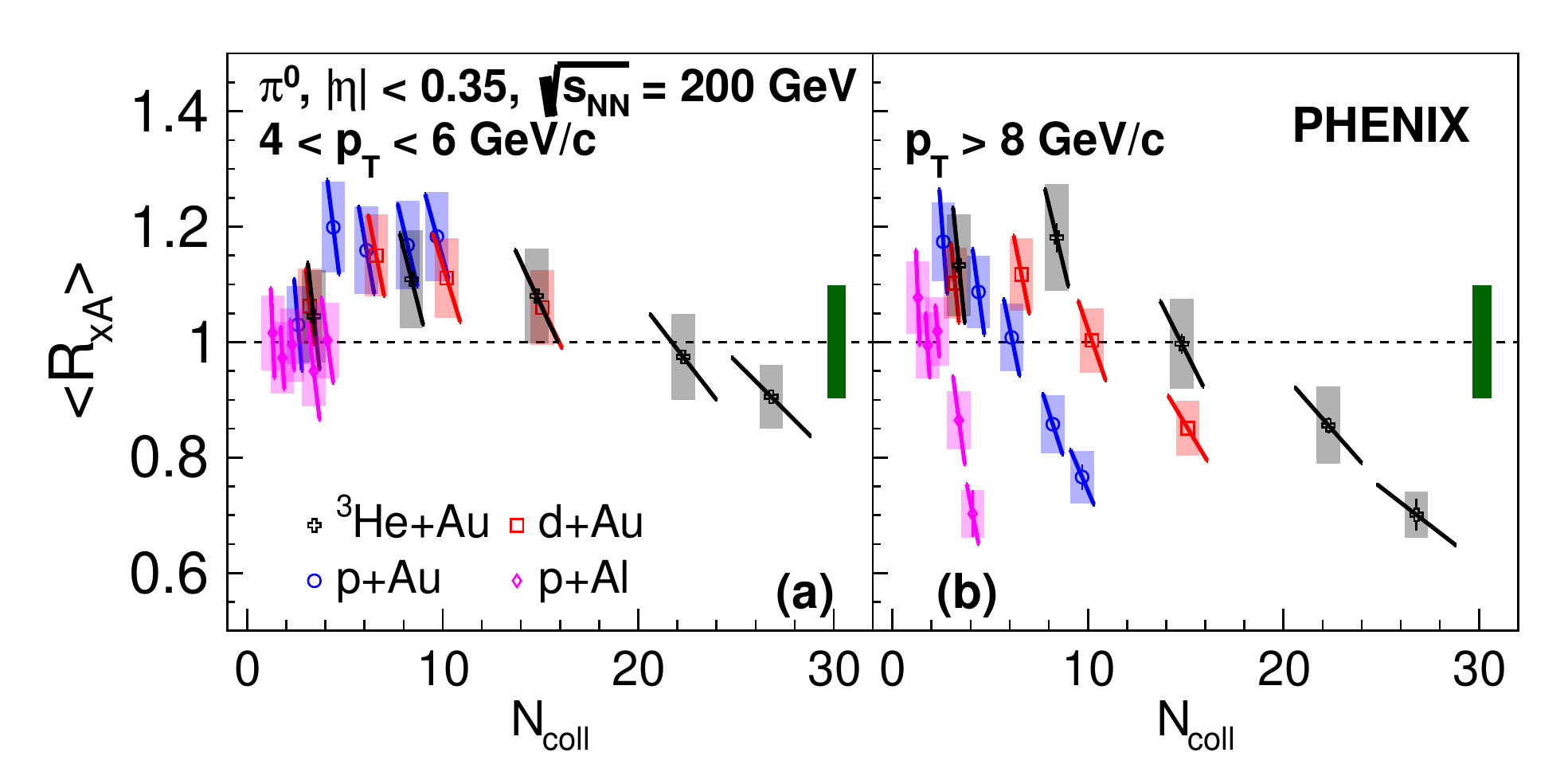}
\includegraphics[width=1.0\linewidth]{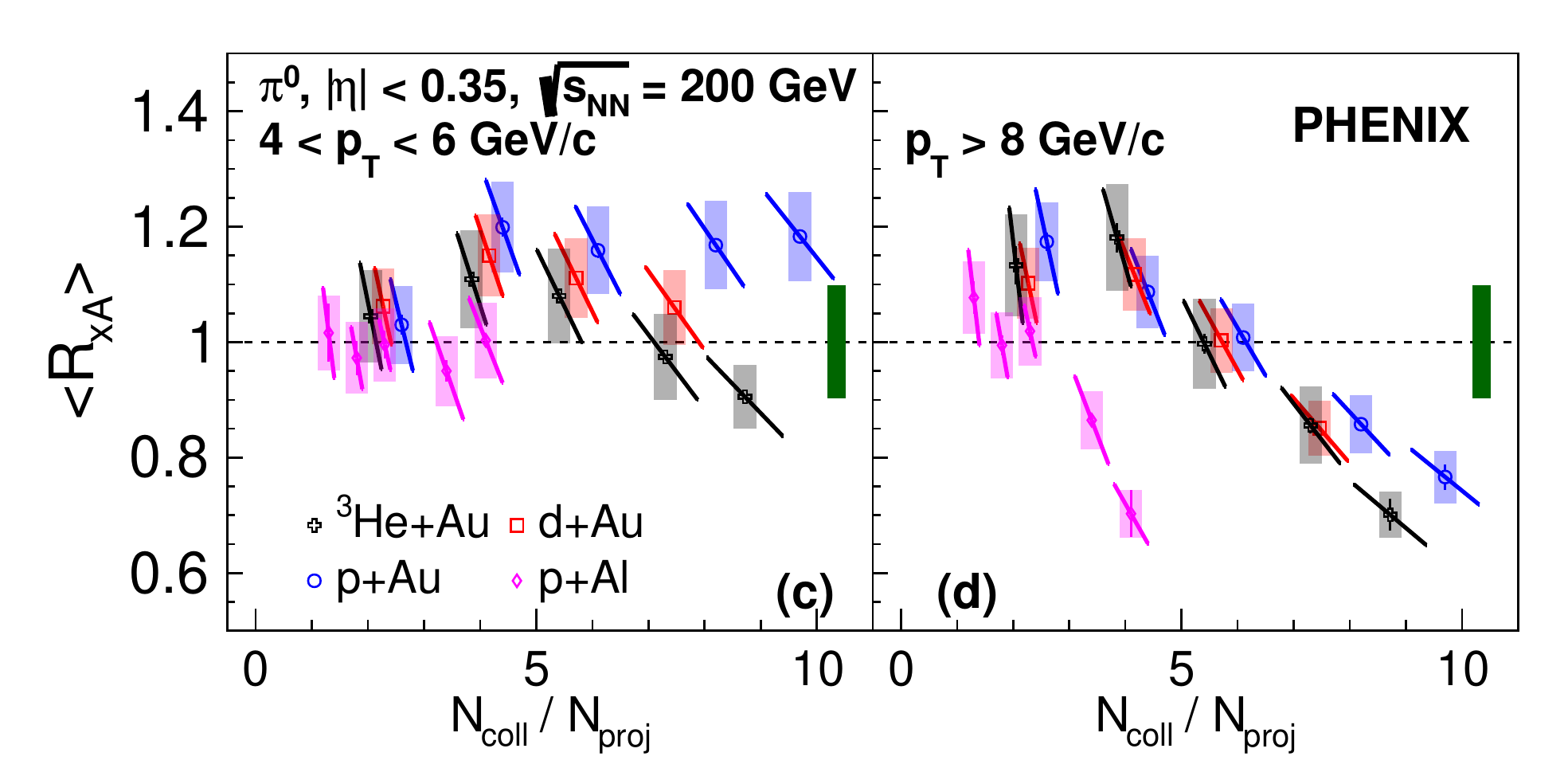}
\caption{Average \raa versus the number of collisions for (a) the region 
around the \raa peak [$4< p_T<6$ GeV/$c$] and (b) the high $p_T$ region 
[$p_T>8$ GeV/$c$]. (c,d) Average \raa versus the number of collisions 
per projectile participant for the same two $p_T$ ranges.  The 
statistical (error bars) and systematic (boxes) uncertainties are 
indicated.  The tilted error bars represent the anti-correlated 
uncertainty on the y and x-axis due to the \Ncoll calculations. The bar 
around unity at the highest $p_T$ shown represents the overall 
normalization uncertainty from \pp collisions.
}
\label{fig:intraa}       
\end{figure}

Figure~\ref{fig:intraa} shows \Araa for the two \pt regions for all 
measured centrality selections from all collision systems. In panels
(a) and (b) \Araa is plotted as function of \Ncoll and in 
panels (c) and (d) as function of \Ncoll per number of participating
nucleons in the projectile \Nproj.

Figure~\ref{fig:intraa}(a) shows \Araa as function of \Ncoll for the 
lower \pt range from 4 to 6 \gevc, covering the peak in \raa for all 
systems. The \Araa is remarkably close to binary scaling, with 
deviations that are visibly smaller than those observed at high \pt 
[see Fig.~\ref{fig:intraa}(b)]. Another notable difference compared to 
the high-\pt range is that all systems show similar deviations from 
binary scaling at the same \Ncoll. In contrast, the systems involving a 
Au target nucleus do not show a common trend with \Ncoll/\Nproj 
[see Fig.~\ref{fig:intraa}(c)] 
These observations are qualitatively the same for any \pt window 
between 1 and 6 \gevc, which suggests that the mechanism underlying the 
nuclear modification is different at high and low \pt with a transition 
in the 5 to 7 \gevc range.

Figure~\ref{fig:intraa}(b) shows that for \pt above 8 \gevc the \Araa 
exhibits no common trend as function of \Ncoll. The \Araa is below 
\Ncoll scaling for central classes and above for peripheral classes 
for all collision systems.  The situation changes when looking \Araa 
versus \Ncoll/\Nproj [see Fig.~\ref{fig:intraa}(d)].  The collision 
systems involving Au as a target nucleus (\pAu, \dAu, and \HeAu) follow 
a common trend. For Al as a target nucleus, a distinctly different 
trend is observed. The modifications to binary scaling or \Araa remain 
approximately the same for similar \pAu and \pAl centrality classes, but 
occur at different \Ncoll/\Nproj. The same trends are observed for any 
choice of \pt threshold above ~7 \gevc up to 15 \gevc, above which the 
statistical precision is limited.  There are two model independent 
conclusions that can be derived from the observations: (i) the 
underlying mechanism for the nuclear modification {does not depend on 
the projectile nucleus}, and (ii) the nuclear modification is not driven 
by the thickness of the nuclear matter traversed by the projectile.

\subsubsection{Model comparison and discussion}

The PDF of a nucleon is modified if the nucleon is within a nucleus and 
the modifications increase with increasing number of nucleons in the 
nucleus.  Similarly to the free proton PDFs themselves, the nuclear 
parton distribution functions (nPDFs) are determined empirically by fitting a 
large variety of experimental data. Here three different nPDFs are 
considered: nNNPDFv1.0~\cite{AbdulKhalek:2019mzd}, 
EPPS16~\cite{Eskola:2016oht}, and nCTEQ15~\cite{Kovarik:2015cma}. 
For consistency, the same framework was used in all 
calculations with the same fragmentation 
function~\cite{deFlorian:2017lwf}.

\begin{figure}[htb]
\includegraphics[width=1.0\linewidth]{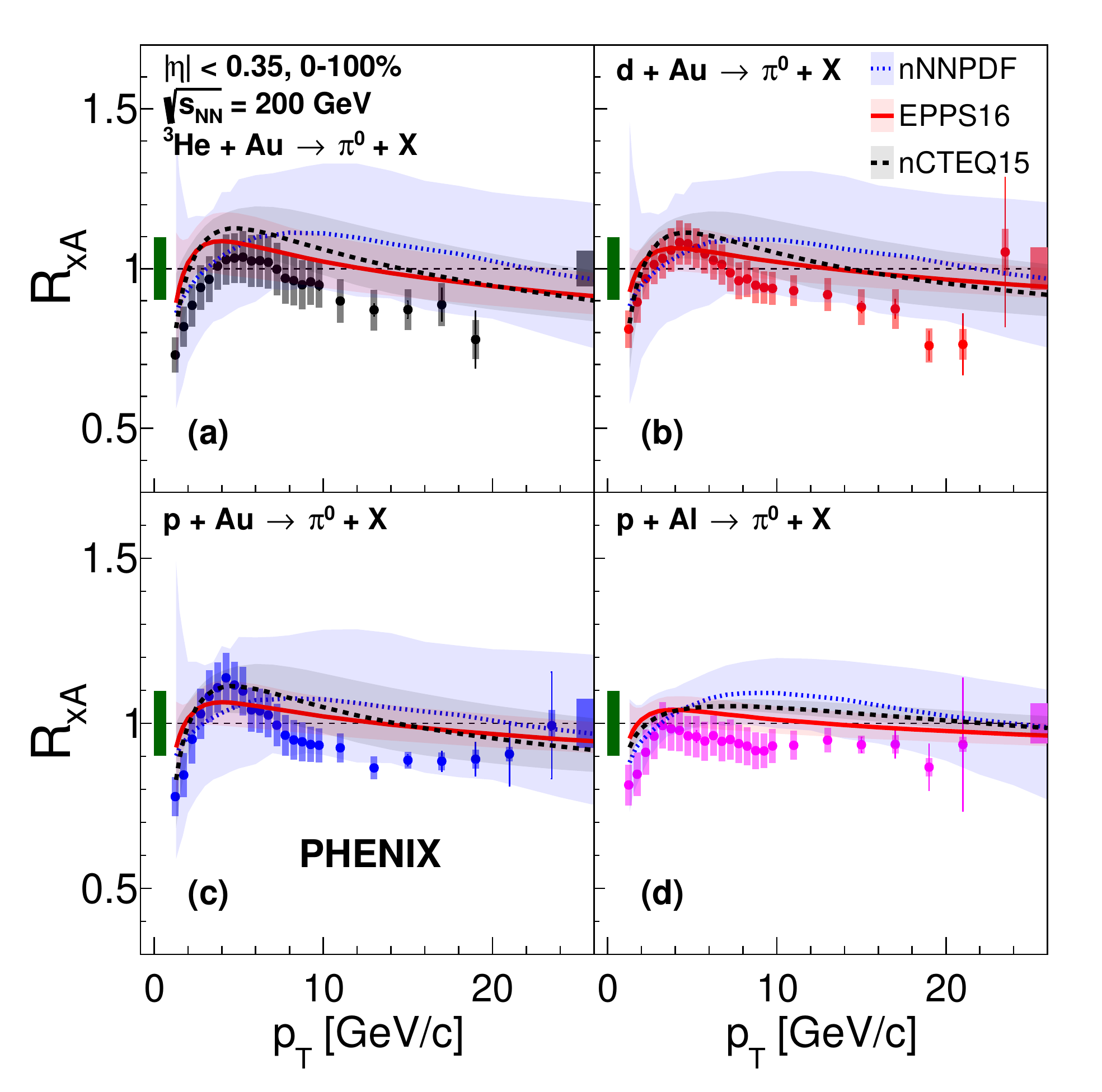}
\caption{\raa for inelastic collisions compared to three different 
nuclear PDF calculations and their uncertainties. The data points 
include the statistical and systematic uncertainties. The left box 
around unity represents the overall normalization uncertainty on the \pp 
collisions and the right box represents the uncertainty from the 
calculated \Ncoll.}
\label{fig:raapdf}       
\end{figure}

\figu{fig:raapdf} compares the measured nuclear-modification factors for 
inclusive \pAl, \pAu, \dAu, and \HeAu collisions are to the predictions 
using the three different nPDFs mentioned above. The central value of 
the predictions is represented by a line and the uncertainties from 
fitting the nPDF to data are given as shaded area.  Due to their large 
uncertainties, all three nPDFs give \raa predictions consistent with the 
data. However, looking at the central values, the predictions are in 
tension with the trends of the data. For example, for the nNNPDF case an 
enhancement is observed from 4 to above 20 \gevc for all systems, with a 
maximum near 8 \gevc, clearly not consistent with data.  Looking at 
individual collision systems, EPPS16 and nCTEQ15 based calculations 
qualitatively, but not quantitatively, capture the general trends.  The 
tension is most clearly visible when comparing the system size 
dependence. Each nPDF calculation predicts an ordering of the 
enhancement of \raa in their respective peak region: 
$\HeAu>\dAu>\pAu>\pAl$, which is significant as the systematic 
uncertainties on the nPDFs within one approach are highly correlated 
between systems. The {predicted} ordering in the lower \pt (2--10\,\gevc) 
region, depending on the model, results from the modification increasing 
both with the target size and with the projectile size. In contrast, the 
data show the reverse ordering $\HeAu<\dAu<\pAu$ with decreasing 
projectile size in the peak region.

For the same reasons that led to predictions of increasing modification 
at lower \pt.  At high-\pt, the models predict an ordering of \raa with 
projectile and target size: $\HeAu<\dAu<\pAu<\pAl$. In contrast, the 
data show a larger suppression than any of the models, and it is 
essentially independent of the collision system. However, given the 
systematic uncertainties on the \raa scale, the nPDF predictions are 
consistent with the data at high \pt. The different trends, in 
particular at low \pt, of the nPDF calculations compared to the data 
suggest that there must be additional physics driving the nuclear 
modification beyond the nPDFs.

\begin{figure}[htb]
\includegraphics[width=1.0\linewidth]{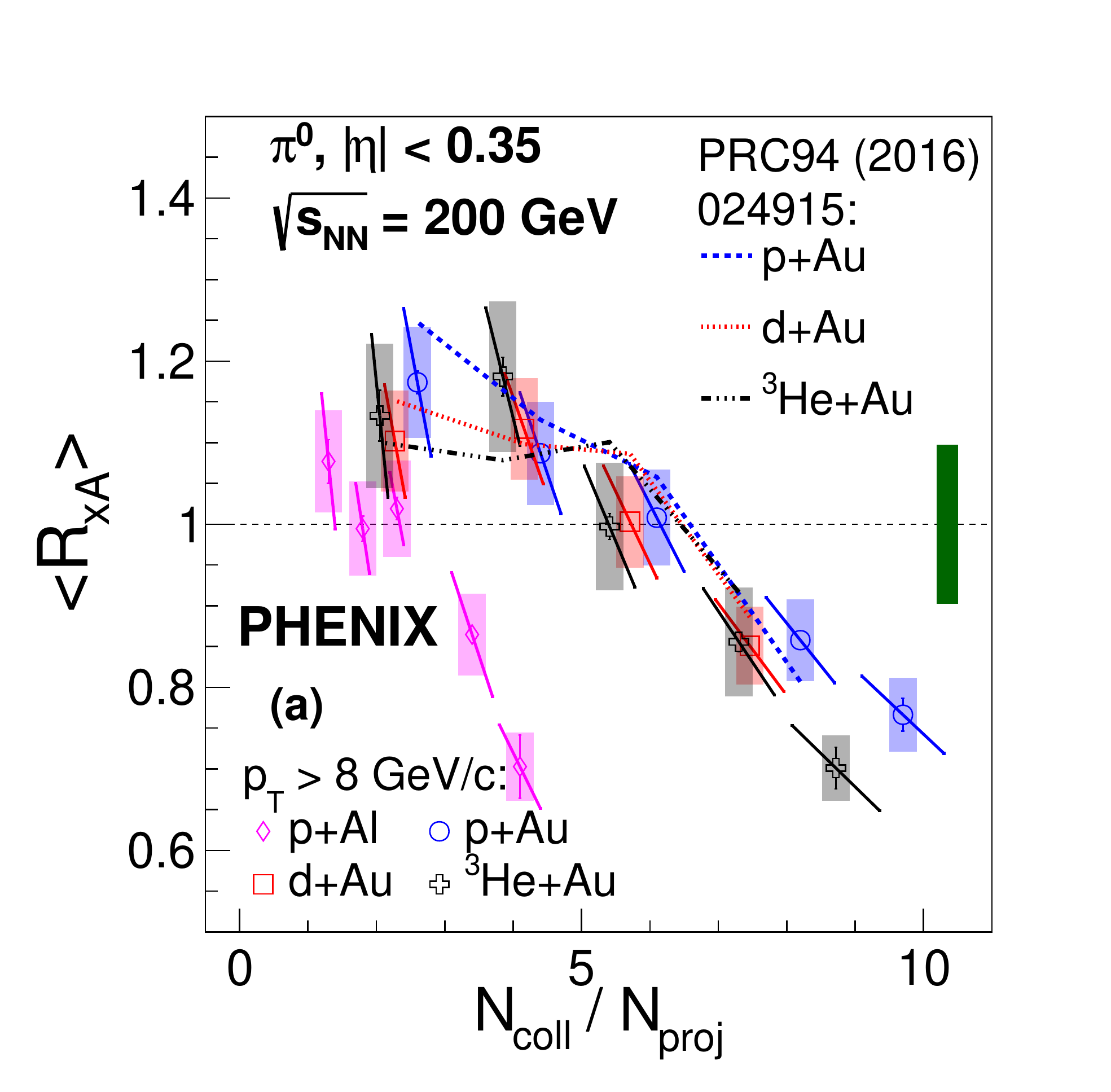}
\includegraphics[width=1.0\linewidth]{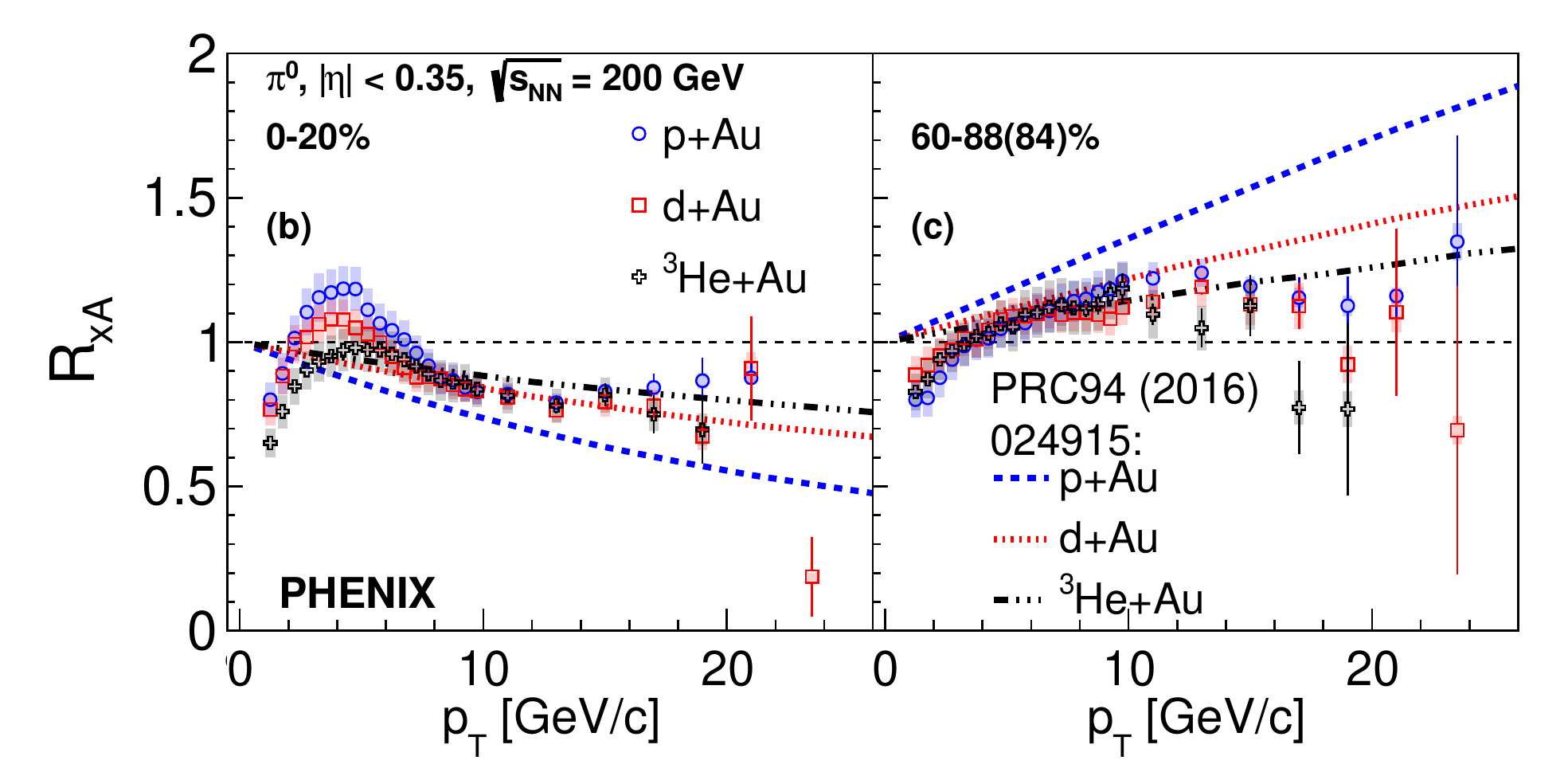}
\caption{(a) The \Araa above $\pt=8$\,\gevc as a 
function \Ncoll/\Nproj with predictions 
from~\cite{McGlinchey:2016ssj} for the consequences of high-$x$ nucleon 
size fluctuations.  (b) The \raa as a function of \pt 
for (b) most-central and (c) most-peripheral collisions. }
\label{fig:raajamie}       
\end{figure}

The data show that at high \pt \piz yields from small systems are 
suppressed relative to \pp in central event selections, while they are 
enhanced for peripheral selections. Furthermore, for \pAu, \dAu, and 
\HeAu, the \Araa values for $\pt>8$\,\gevc are consistent with a 
superposition of independent collisions of the projectile nucleons. At 
the same time, \pAu and \pAl show nearly the same \Araa in the same 
centrality bin selection. These observations contradict any scenarios 
where parton energy loss would be responsible for the modification, 
which would necessarily result in an ordering of \raa values as $\HeAu < 
\dAu < \pAu < \pAl \leq 1$ for the system dependence, with the 
suppression for each system being largest for central and $\raa\approx1$ 
for peripheral collisions.

Models that invoke nucleon size variations have been proposed to explain 
the suppression/enhancement pattern seen in the 
data~\cite{Alvioli:2014eda, Alvioli:2017wou}. These models assume that 
nucleons with high-$x$ partons have a more compact color configuration 
and thus will produce on average less binary collisions and target 
participants at the same impact parameter as nucleons without high-$x$ 
partons. As a consequence, events with a high \pt \piz would typically 
be biased towards smaller multiplicity of the overall event, leading to 
an apparent enhancement in peripheral event selections and a suppression 
in central events. The calculations from~\cite{McGlinchey:2016ssj}, 
which predicted jet \raa for \pAu and \HeAu based on a comparison to 
\dAu data\footnote{Note that jet \raa presented 
in~\cite{McGlinchey:2016ssj} was converted to \piz \raa assuming 
$\pt(\piz)=0.7\,\pt^{\rm jet}=0.7 \times 100\,{\rm GeV} \times x_p$ and 
$\Araa\approx\raa(\pt)$. This procedure was discussed with the authors.}, 
are compared to \piz \Araa above {a \pt of} 8\,\gevc , 
[see \fig{fig:raajamie}(a)]. The observed centrality dependence is 
quite consistent with the data. It can be expected that in this model 
the same event selection bias would occur in \pAl collisions.

Although this model plausibly describes the \dAu and \HeAu data, it 
particularly misses the \pAu. Additionally, it is important to note that 
this model predicts an ordering of \raa with system size and centrality 
at higher \pt. \figu{fig:raajamie} clearly shows that for (b) central 
collisions the predicted \raa values follow $\HeAu>\dAu>\pAu$ and 
for (c) peripheral collisions the ordering is reversed.  In contrast, such 
an ordering is not supported by the data.

In Ref.~\cite{Kordell:2016njg}, the bias of the event selection by 
centrality occurs because soft particle production away from the hard 
scattering process is suppressed, caused by the depletion of energy 
available in the projectile after the hard scattering process. The \raa 
calculated for \dAu with this model was consistent with 
preliminary~\cite{Kordell:2016njg} and final \dAu data within systematic 
uncertainties. It would be interesting to see these calculations 
expanded to the full variety of available data from small systems.

\begin{figure}[htb]
\includegraphics[width=0.95\linewidth]{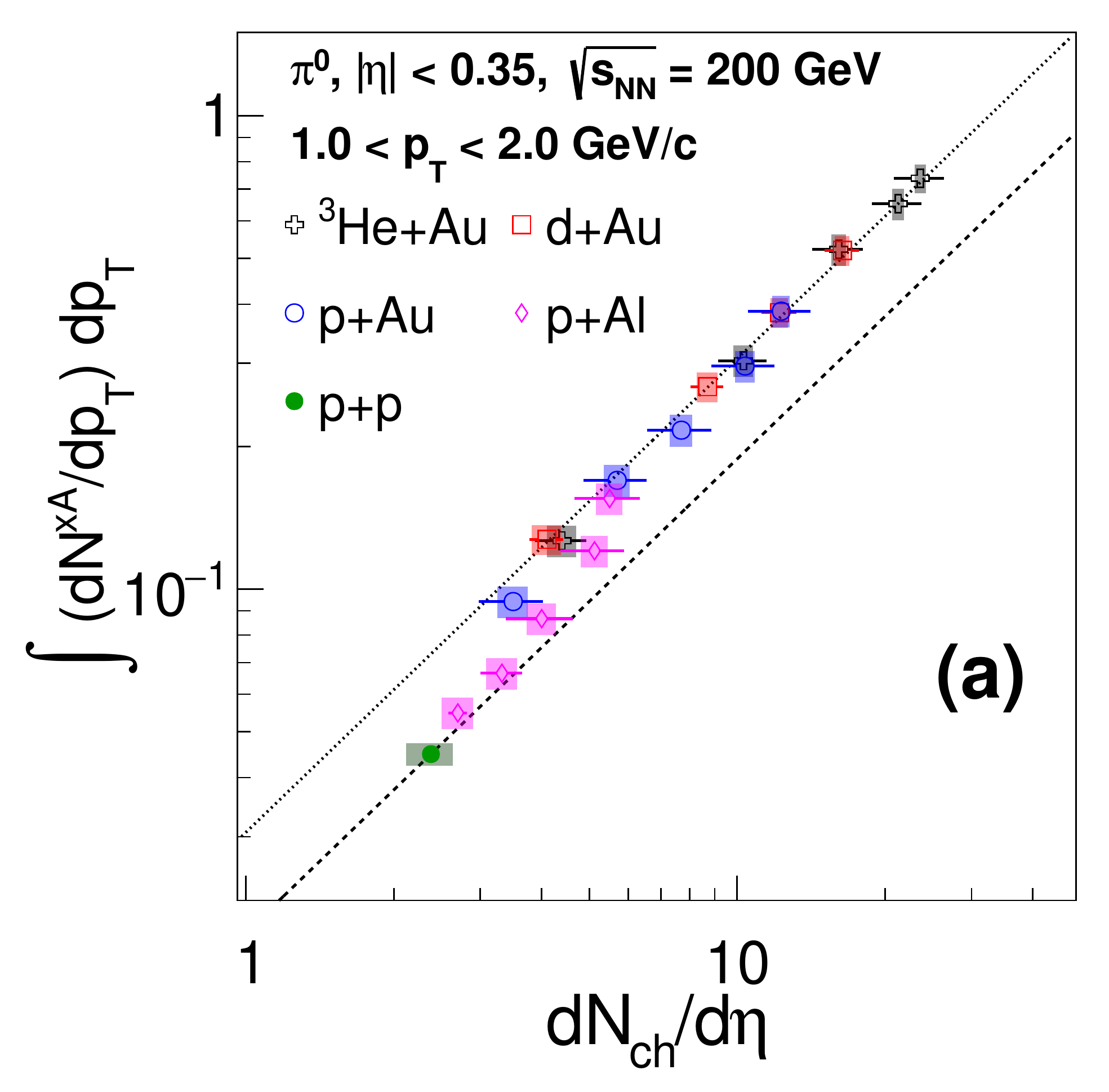}
\includegraphics[width=0.95\linewidth]{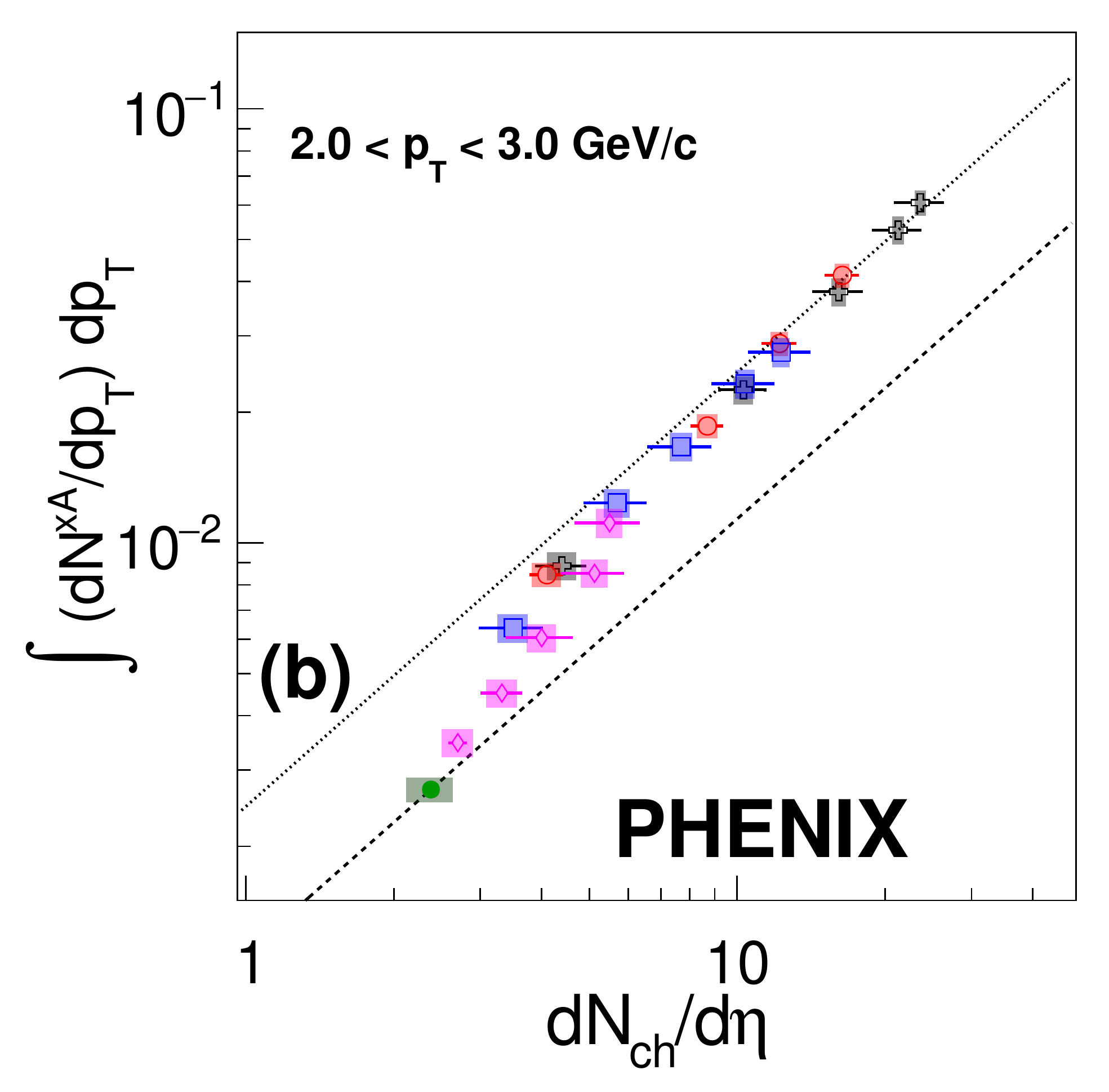}
\caption{Integrated yields for (a) 1--2 \gevc and (b) 2--3 \gevc 
as a function of charged particle multiplicity density at 
midrapidity.  The lines are explained in the text.
 }
\label{fig:Abaa}       
\end{figure}

In recent years particle spectra from \pp collisions at the LHC have 
been interpreted in the context of hydrodynamic models and the presence 
of strong radial 
flow~\cite{Bhattacharyya:2015hya,Sarkar:2016ikv,Hirono:2017svi,Khuntia:2018znt}, 
but no predictions exist for small systems at RHIC energies that could 
be compared to the data. If the large anisotropies of particle 
production seen at RHIC in \pAu, \dAu, and \HeAu are indeed related to 
hydrodynamic expansion of the collision volume, as suggested 
in~\cite{PHENIX:2018lia}, then the same systems must also exhibit radial 
flow because the anisotropy would be a geometry driven modulation of 
radial flow. The effects of radial flow are typically most prominent at 
\pt below a few \gevc, where soft particle production mechanisms 
dominate. In the presence of radial flow the \piz yield would be shifted 
towards higher momentum by the velocity field. Accordingly, when 
comparing the shape of the \piz momentum spectra from \xA to that from 
\pp, a depletion of the yield at the lowest \pt is expected, while at 
higher \pt the yield would be enhanced with a transition near the \piz 
mass. Because the \pt range of the \piz data starts at 1~\gevc, only the 
region where an enhancement would be expected can be studied here.

To look for possible indications of radial flow the integrated yields 
are calculated for two \pt ranges, 1--2 and 2--3 \gevc, for all systems 
and event selections. The results are plotted in \fig{fig:Abaa} as 
functions of the charged particle multiplicity density $dN_{\rm ch}/d\eta$ at 
midrapidity for the corresponding system and event selection. Also 
shown on each panel are two lines indicating integrated yields {linearly 
increasing with} $dN_{\rm ch}/d\eta$. The lower line is anchored to the \pp 
point following a trend of unchanged shape of the spectra, and the upper 
line matches the yield for the 0\%--20\% \HeAu selection. While the 
peripheral \pAl events follow the \pp trend, all other selections show 
higher integrated yields compared to the \pp trend. Above 
$dN_{\rm ch}/d\eta \approx 10$ the data tends to be proportional to 
$dN_{\rm ch}/d\eta$ again but at a much higher level.

The observed trend is qualitatively consistent with the presence of 
radial flow in small systems. Interestingly, the surprisingly rapid 
transition over the {$dN_{\rm ch}/d\eta$} range from $\approx$~3~to~10 is 
similar to recent observations of low \pt direct photon 
emission~\cite{ppg:212} {and strangeness production~\cite{ALICE:2017jyt}. 
Both also indicate} a transition from \pp-like emission to a significant 
enhancement at similar event multiplicities. Furthermore, the presence 
of radial flow could naturally explain the much larger observed Cronin 
effect for protons from small systems~\cite{Antreasyan:1978cw}, which so 
far has eluded a quantitative understanding. However, before drawing 
firm conclusions, more investigations are necessary. These should 
include the study of heavier hadrons, such as Kaons and protons.

\section{Summary}

In summary, this paper presents new measurements of the invariant cross 
section of neutral pion production from \pp collisions and invariant 
yields from \pAl, \pAu, \dAu, and \HeAu at $\sqrt{s_{_{NN}}} = 200\,\gev$. 
For \pp the new results extend the measured range {up} to $\pt \approx 
25\,\gev/c$ and improve statistical and systematic uncertainties 
compared to the previous measurement. NLO pQCD calculations are found to 
be consistent with the data as previously reported. For \pAl, \pAu, 
\dAu, and \HeAu collisions at $\sqsn = 200\,\gev$, \piz \pt spectra from 
inelastic collisions and from centrality selected event samples were 
measured, including a sample of the 0\%--5\% most central events for 
\pAl, \pAu, and \HeAu, which was recorded with a dedicated high 
multiplicity trigger.

At high transverse momentum ($\pt>8~\gevc$), where hard scattering 
processes are the dominant production mechanism, the nuclear 
modification factors for all collision systems are found to be nearly 
constant. For the same event selection in percent centrality, different 
collision systems exhibit {a} value of \raa that is compatible within 
uncertainties. For the full inelastic cross section, \raa is consistent 
with unity, pointing towards little or no nuclear modification of hard 
scattering processes in small systems. For the most central events, it 
is observed that \raa is significantly below unity. However, \raa 
increases monotonically with decreasing centrality and exceeds unity for 
peripheral collisions. For Au target nuclei, the \Araa above \pt of 8 
\gevc shows a common trend with \Ncoll/\Nproj.  This indicates that, for 
hard scattering processes, the nucleons in the small projectile nucleus 
interact mostly independently with the Au target. For \pAl collisions, 
\Araa does not follow the same trend.  At the same event centrality, the 
\pAl \Araa is the same as for \pAu, which suggests that the mechanism 
that causes the change of \raa with centrality does not depend on the 
target nucleus.

These observations disfavor scenarios where energy loss is a significant 
contributor to the nuclear modification of high \pt particle production 
in small systems. The counter-intuitive centrality dependence is likely 
linked to a mismatch of the centrality selection of events using charged 
particle multiplicity and mapping them to a number of binary collisions 
using the standard Glauber model. {In this picture,} events with a high 
\pt \piz are biased towards smaller underlying event multiplicity.  This 
might be due to physical fluctuations of the proton size or simply due 
to energy conservation if high \pt jets are present.

For lower \pt, \raa for all systems initially increases with \pt and 
reaches a peak near 4--6\,\gevc for central and semi-central 
collisions. For peripheral collisions, \raa levels off to a constant at 
approximately the same high \pt value for all systems. For inelastic 
collisions and more central collisions, \raa resembles what has been 
referred to as the Cronin effect in fixed target experiments - a rise, 
followed by a peak, followed by a plateau. However, unlike at lower 
energies, \pp and \AB \piz cross sections are not related by a power 
$(xA)^{n(\pt)}$ with a common $n(\pt)$. Furthermore, the peak height 
value around 4--6\,\gevc shows a clear system size dependence 
$\pAu>\dAu>\HeAu>\pAl$, where the \raa peak height value is well above 
unity for \pAu and drops to close to unity for \pAl collisions.

While the shape of \raa roughly resembles what is expected from the 
nuclear modification of PDFs, the observed system size dependence of the 
peak height of \raa shows exactly the reverse ordering of what was 
predicted by nPDF calculations. Therefore it is likely that nPDFs 
{alone} are insufficient to explain the nuclear modifications in small 
systems.

In the same \pt region, \Araa was used to study the dependence on 
centrality. For all systems, \Araa in the range 4--6\,\gevc follows a 
common trend with \Ncoll. At high \pt, \Araa scales with \Ncoll/\Nproj 
for Au target nuclei.  While at lower \pt, \dAu and \HeAu are not a 
superposition of \pAu-like collisions. Consequently, different 
mechanisms must contribute to the nuclear modification at high and low 
\pt. For high $p_T$, the apparent centrality dependence is likely 
due to a bias in the event selection. At lower $p_T$, final state effects 
related to the presence of interacting hadrons may be at play. If a QGP 
droplet is indeed produced during the collision, radial flow may be 
one possible mechanism to explain this trend, although 
further investigation is needed that is outside the scope of this paper.


\begin{acknowledgments}

We thank the staff of the Collider-Accelerator and Physics
Departments at Brookhaven National Laboratory and the staff of
the other PHENIX participating institutions for their vital
contributions.  
We also thank I. Helenius and J. Rojo for the nPDF calculations 
plus M. van Leeuwen for the NLO calculations.
We acknowledge support from the Office of Nuclear Physics in the
Office of Science of the Department of Energy,
the National Science Foundation,
Abilene Christian University Research Council,
Research Foundation of SUNY, and
Dean of the College of Arts and Sciences, Vanderbilt University
(U.S.A),
Ministry of Education, Culture, Sports, Science, and Technology
and the Japan Society for the Promotion of Science (Japan),
Conselho Nacional de Desenvolvimento Cient\'{\i}fico e
Tecnol{\'o}gico and Funda\c c{\~a}o de Amparo {\`a} Pesquisa do
Estado de S{\~a}o Paulo (Brazil),
Natural Science Foundation of China (People's Republic of China),
Croatian Science Foundation and
Ministry of Science and Education (Croatia),
Ministry of Education, Youth and Sports (Czech Republic),
Centre National de la Recherche Scientifique, Commissariat
{\`a} l'{\'E}nergie Atomique, and Institut National de Physique
Nucl{\'e}aire et de Physique des Particules (France),
Bundesministerium f\"ur Bildung und Forschung, Deutscher Akademischer
Austausch Dienst, and Alexander von Humboldt Stiftung (Germany),
J. Bolyai Research Scholarship, EFOP, the New National Excellence
Program ({\'U}NKP), NKFIH, and OTKA (Hungary),
Department of Atomic Energy and Department of Science and Technology
(India),
Israel Science Foundation (Israel),
Basic Science Research and SRC(CENuM) Programs through NRF
funded by the Ministry of Education and the Ministry of
Science and ICT (Korea).
Physics Department, Lahore University of Management Sciences (Pakistan),
Ministry of Education and Science, Russian Academy of Sciences,
Federal Agency of Atomic Energy (Russia),
VR and Wallenberg Foundation (Sweden),
the U.S. Civilian Research and Development Foundation for the
Independent States of the Former Soviet Union,
the Hungarian American Enterprise Scholarship Fund,
the US-Hungarian Fulbright Foundation,
and the US-Israel Binational Science Foundation.

\end{acknowledgments}



%
 
\end{document}